\documentclass[9pt,twocolumn,twoside]{pnas-new}

\templatetype{pnasresearcharticle}

\usepackage{ulem}
\usepackage{xcolor}

\usepackage{hyperref}
\hypersetup{colorlinks,allcolors=blue}

\title{Stem-loop formation drives RNA folding in mechanical unzipping experiments}

\author[a]{Paolo Rissone}
\author[b]{Cristiano V. Bizarro} 
\author[a,1]{Felix Ritort}

\affil[a]{Small Biosystems Lab, Condensed Matter Physics Department, University of Barcelona, Barcelona, 08028, Spain}
\affil[b]{Instituto Nacional de Ciência e Tecnologia em Tuberculose, Centro de Pesquisas em Biologia Molecular e Funcional, Pontifícia Universidade Católica do Rio Grande do Sul, 90616-900, Porto Alegre, Rio Grande do Sul, Brazil}

\leadauthor{Rissone}

\authorcontributions{C.V.B and F.R. designed research; P.R. analyzed and modelled data; C.V.B. prepared samples and did the experiments; P.R., C.V.B and F.R. wrote the paper.}
\authordeclaration{The authors declare no conflict of interest.}
\correspondingauthor{\textsuperscript{1}To whom correspondence should be addressed. E-mail: ritort@ub.edu}

\begin{abstract}
Accurate knowledge of RNA hybridization is essential for understanding RNA structure and function. Here we mechanically unzip and rezip a 2kbp RNA hairpin and derive the ten nearest-neighbor base-pair (NNBP) RNA free energies in sodium and magnesium with 0.1 kcal/mol precision using optical tweezers. Notably, force-distance curves (FDCs) exhibit strong irreversible effects with hysteresis and several intermediates, precluding the extraction of the NNBP energies with currently available methods. The combination of a suitable RNA synthesis with a tailored pulling protocol allowed us to obtain the fully reversible FDCs necessary to derive the NNBP energies. We demonstrate the equivalence of sodium and magnesium free-energy salt corrections at the level of individual NNBP. To characterize the irreversibility of the unzipping-rezipping process, we introduce a barrier energy landscape of the stem-loop structures forming along the complementary strands, which compete against the formation of the native hairpin. This landscape correlates with the hysteresis observed along the FDCs. RNA sequence analysis shows that base stacking and base-pairing stabilize the stem-loops that kinetically trap the long-lived intermediates observed in the FDC. Stem-loops formation appears as a general mechanism to explain a wide range of behaviors observed in RNA folding.
\end{abstract}

\begin{document}

\maketitle
\thispagestyle{firststyle}
\ifthenelse{\boolean{shortarticle}}{\ifthenelse{\boolean{singlecolumn}}{\abscontentformatted}{\abscontent}}{}

\dropcap{U}nzipping experiments permit to investigate the physico-chemical properties of nucleic acids, from the thermodynamics of duplex formation to the folding of secondary and tertiary structures. In particular, DNA hybridization finds diverse applications in the field of DNA nanotechnology,  the construction of DNA origami, molecular robots, DNA walkers, switches and nanomotors \cite{castro2011primer,cha2014synthetic,hagiya2014molecular,wang2015dna,jung2016stochastic}. In an unzipping experiment, the two strands of a duplex DNA or RNA molecule are mechanically pulled apart by exerting opposite forces on the two strands on one end. In this way, it is possible to measure a force-distance curve (FDC) that exhibits a sequence-dependent sawtooth pattern. DNA unzipping has been used to test the validity of the nearest-neighbor (NN) model \cite{devoe1962stability,crothers1964theory,breslauer1986predicting,santalucia1998unified} and to extract the ten NN base-pairs (NNBP) free-energy parameters at different salt conditions \cite{huguet2010single, huguet2017derivation}. A precise knowledge of the NNBP energies might be also useful to unravel hidden energy codes in molecular evolution \cite{klump2020energy}. 

Here we derive the ten NNBP RNA energies from unzipping experiments carried out on a 2kbp RNA hairpin in monovalent (sodium) and divalent (magnesium) salt conditions. The NN model has many parameters requiring a sufficiently long RNA hairpin to infer them from unzipping experiments reliably. Two are the main difficulties of these experiments: first, the molecular synthesis of a long (a few kb) RNA hairpin is challenging; second, the FDC along the RNA sequence alternates reversible unzipping regions with irreversible ones that exhibit hysteresis, and multiple long-lived intermediates  \cite{chen2000rna,zhuang2003single}. Compared to DNA, where unzipping is practically reversible, a similar derivation of the RNA energies from irreversible FDCs is not possible. Here we derive the full equilibrium FDC in RNA by the piecewise assembly of the reversible parts and the reconstructed equilibrium ones for the irreversible regions. These are obtained by repeatedly unzipping and rezipping the RNA hairpin in these irreversible regions and using statistical physics methods based on fluctuation theorems. This allows us to derive the NNBP energies for RNA in sodium and magnesium and compare them with the results reported by the literature \cite{mathews1999expanded, walter1994coaxial, xia1998thermodynamic, freier1986improved}. Moreover, we demonstrate the validity of an equivalence rule for the free-energy salt corrections between sodium and magnesium at the level of individual NNBP. We find that NNBP free-energy parameters for a given magnesium concentration are equal to those in $77(\pm49)$-fold sodium. This result is compatible with the 100/1 rule of thumb by which the non-specific RNA binding affinity of 10mM Mg$^{2+}$ approximately equals that of 1M Na$^{+}$\cite{bizarro2012non}. We provide a solid verification of this phenomenological result by measuring the NNBP RNA energies in sodium and magnesium. We study the irreversibility and hysteresis in the FDCs and hypothesize that this is caused by the formation of stem-loop structures along the unpaired single strands. Remarkably, the hysteresis along the unzipping-rezipping pathway directly correlates with the barrier energy landscape defined by the stem-loops that are formed at the junction separating single strands and duplex. A sequence analysis of the irreversible regions of the 2kbp RNA and experiments on specifically designed short-RNA sequences demonstrates that base stacking and base-pairing within the ssRNA promote the formation of stem-loop RNA structures transiently stabilized at forces as high as 20pN. The stem-loops mechanism explains the slow kinetics and multiple trapping conformations observed in RNA folding, with implications for the RNA folding problem \cite{Bryngelson1995,chen2000rna,woodson2010compact,ferreiro2014frustration,englander2017case}.
%
%
\section*{Results}
\begin{figure*}[ht]
\centering
\includegraphics[width=.85\textwidth]{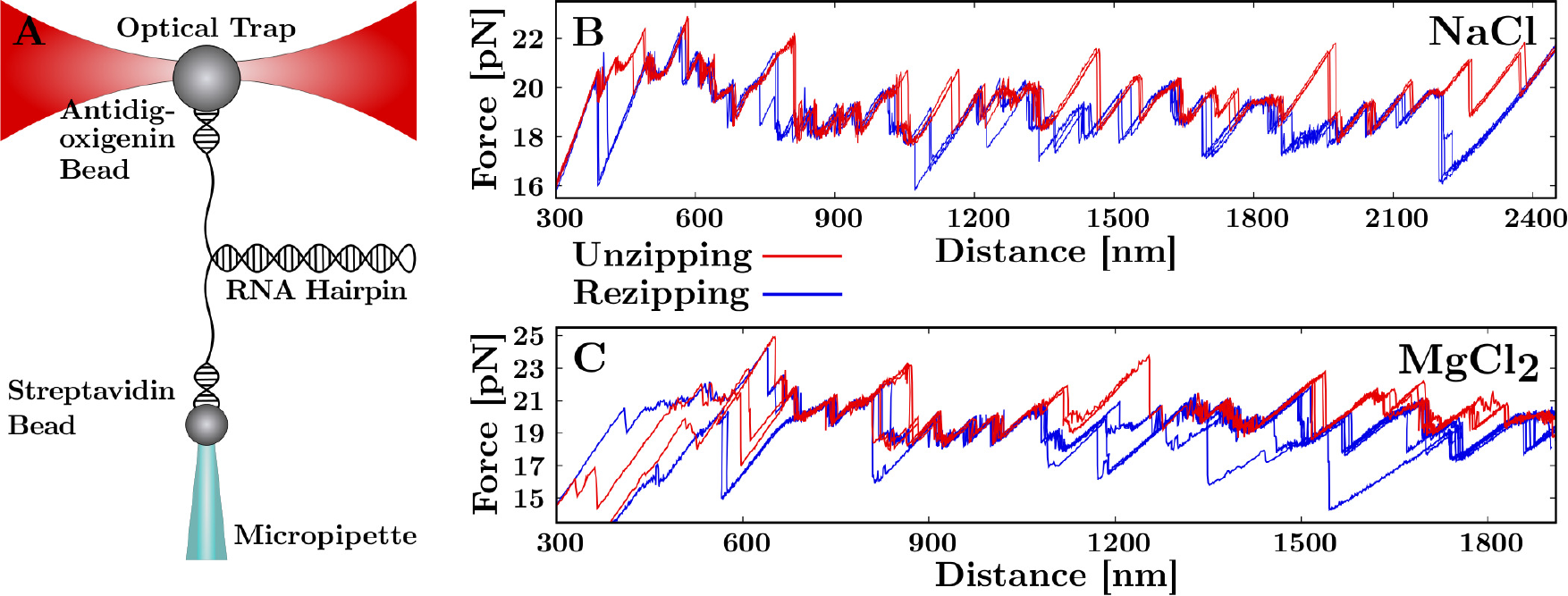}
\caption{\label{fig:exp_setup} Experimental setup and measured FDCs in sodium and magnesium. (\textbf{A}) Optical tweezers setup. The RNA hairpin is mechanically unzipped and rezipped by moving the optical trap. (\textbf{B}) Unzipping/rezipping FDCs (red/blue) in 500mM NaCl. Hysteresis is apparent in some regions of the FDC. (\textbf{C}) Unzipping/rezipping FDCs in 10mM MgCl$_2$. Magnesium enhances the amount of hysteresis as compared to the sodium case. The irreversibility is so large that the initial and final regions of the FDCs remain inaccessible.}
\end{figure*}

We used optical tweezers to pull a 2027bp RNA hairpin with short (29bp) hybrid DNA/RNA handles. Details on the hairpin and the synthesis protocol are given in Material and Methods. In our setup, digoxigenin (DIG)-labeled and biotin-labeled handles of the hairpin are connected to anti-DIG (AD) and streptavidin-coated (SA) beads, respectively. The AD bead is optically trapped while the SA bead is immobilized by air suction at the tip of a micropipette (Fig.\ref{fig:exp_setup}A). By moving the optical trap upwards, the molecule gradually unzips from the completely folded double-stranded RNA (dsRNA) hairpin conformation (the native state $N$) to the completely unfolded and stretched single-stranded RNA (ssRNA) conformation (the unfolded state $U$) producing the characteristic sawtooth pattern of the FDC (red curves in Fig.\ref{fig:exp_setup}B,C). Once the hairpin is unfolded the reverse process (rezipping) starts: the trap is moved in the opposite direction and the molecule gradually refolds into the native stem (blue curves in Fig.\ref{fig:exp_setup}B,C).

\begin{figure*}[ht]
\centering
\includegraphics[width=.85\textwidth]{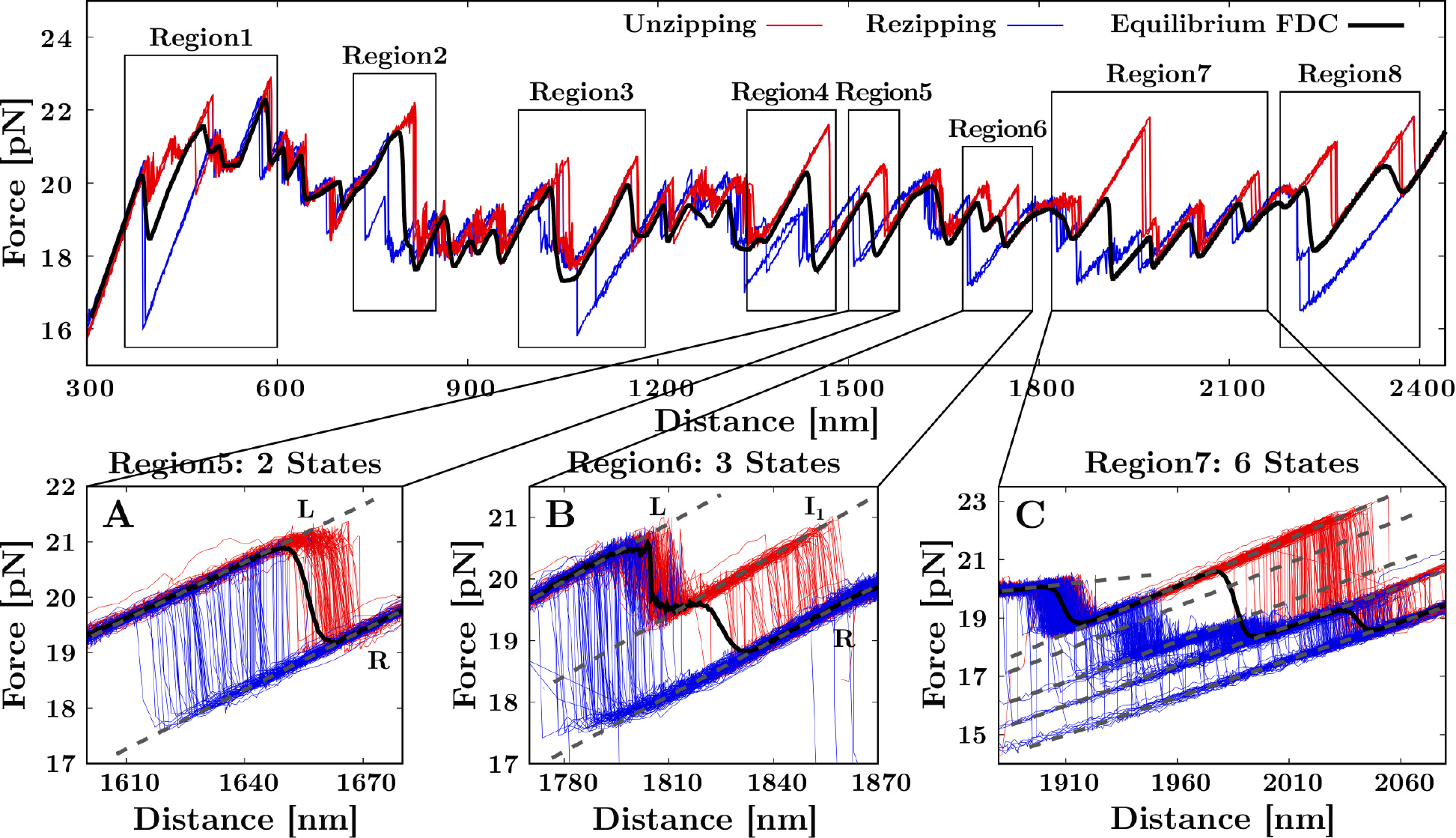}
\caption{\label{fig:NaCl} Unzipping/rezipping FDCs (red/blue) in 500mM NaCl. Black frames mark the irreversible regions. The insets show repeated pulling cycles in regions of increasing complexity. The intermediates (dashed grey lines) and the recovered equilibrium FDC (solid black line) are shown. $(\textbf{A})$ shows a 2-states region ($L/R$) with no intermediates, whereas $(\textbf{B})$ and $(\textbf{C})$ report a 3-states region ($L$,$R$ and the intermediate $I_1$) and a 6-states region ($L$,$R$ and intermediates $I_p$ with $p=1,\ldots,4$), respectively. The equilibrium FDC in the main box (black line) results by merging the reversible FDCs obtained for each region.}
\end{figure*}

\begin{figure*}[ht]
\centering
\includegraphics[width=.85\textwidth]{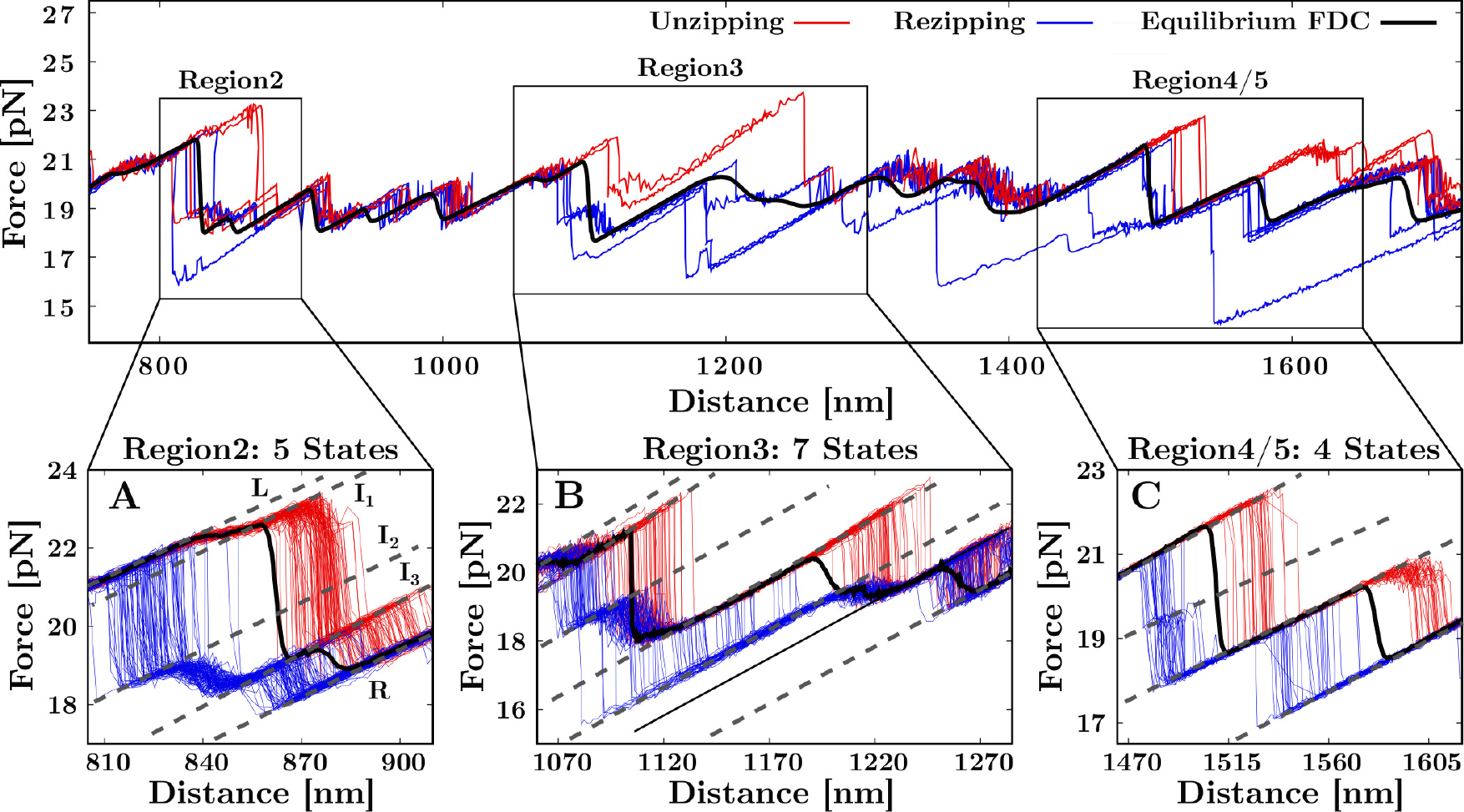}
\caption{\label{fig:MgCl2} Unzipping/rezipping FDCs (red/blue) in 10mM $\rm MgCl_2$. Black frames mark the irreversible regions along the sequence. The insets show repeated pulling cycles in regions of different complexity. Intermediates (dashed grey lines) and the recovered equilibrium FDC (solid black line) are shown. $(\textbf{A})$ shows a 5-states region ($L$,$R$ and intermediates $I_1,I_2,I_3$),  $(\textbf{B})$ shows a 7-states region ($L$,$R$ and intermediates $I_p$ with $p=1,\ldots,5$) and $(\textbf{C})$ shows a 4-states region ($L$,$R$ and intermediates $I_1,I_2$). The equilibrium FDC in the main box (black line) results by merging the reversible FDCs obtained for each region.}
\end{figure*}

The experiments have been performed in buffers containing 100 mM tris(hydroxymethyl)aminomethane (Tris)-HCl (pH 8.1), 1 mM ethylenediaminetetraacetic, and 500 mM NaCl (monovalent salt) or 100mM Tris-HCl (pH 8.1) and 10mM $\rm MgCl_2$ (divalent salt). Notice that the ionic strength of the buffers has to be corrected by adding $\rm 100mM$ Tris-HCl $\rm \equiv 52mM \, [Mon^+]$. Measured FDCs show that changing from [Na$^+$] to [Mg$^{++}$] strongly increases the irreversibility and hysteresis of the FDC. This makes the beginning (the first 200bp, between 400nm and 650nm) and the end (the last 600bp, between 1800nm and 2200nm) of the FDC to become experimentally inaccessible: the RNA hairpin does not hybridize in the experimental timescale (Fig.\ref{fig:exp_setup}C). The observed hysteresis occurs in correspondence of specific regions along the FDC, each one limited by the equilibrated left ($L$) and right ($R$) states, and exhibiting intermediate states $I_p$, with $p=1,\dots,P$. To efficiently sample the intermediates, we repeatedly unzipped and rezipped the RNA between the two limit positions $(L,R)$ typically collecting a hundred trajectories per region. We have identified 8 irreversible regions in sodium (Fig.\ref{fig:NaCl}) and 3 in magnesium (Fig.\ref{fig:MgCl2}). Regions in sodium are numbered from 1 to 8. In magnesium, regions are numbered from 2 to 4/5 to underline the matching of the RNA sequences in those regions in sodium and magnesium, as evidenced by the number of opened base pairs. The larger hysteresis observed in magnesium makes regions 4 and 5 in sodium merge into a single irreversible region (4/5). The missing regions (1 and 6-8) in magnesium results from their inaccessibility, as explained above. Although a few regions do not contain intermediates (e.g. region 5 in Fig.\ref{fig:NaCl}A), most of them exhibit more than one. The level of complexity of the unzipping-rezipping FDCs can be high, e.g. region 3 in magnesium shows 7 states (5 intermediates plus L and R, Fig.\ref{fig:MgCl2}B).

To derive the NNBP energies we computed the equilibrium FDC by applying the extended fluctuation relation (Materials and Methods), which has been introduced to recover the free energy of thermodynamic branches \cite{junier2009recovery}, kinetic states \cite{alemany2012experimental} and ligand binding energies \cite{camunas2017experimental,sonar2020effects}. This allowed us to reconstruct the equilibrium FDCs (black line in Figs.\ref{fig:NaCl} and \ref{fig:MgCl2}) for 7 molecules in sodium and 4 molecules in magnesium.

\subsection*{Derivation of the NNBP energies for RNA}
In the NN model the free energy of formation $\Delta G_0$ of a DNA and RNA duplex is defined as the sum over all adjacent NNBP along the sequence, $\Delta G_0=\sum_i{\Delta g_{0,i}}$ with $\Delta g_{0,i}$ the free energy of NNBP motif $i$. There are 16 different NNBP which energies are degenerated due to Watson-Crick complementarity, reducing the free energies set ($\Delta g_{0,i}$) to 10 parameters. The NNBP energies have been extracted from melting experiments of short RNA duplexes of varying sequence and length \cite{mathews1999expanded, walter1994coaxial, xia1998thermodynamic, freier1986improved}. These values are accessible in the Mfold server \cite{zuker2003mfold}. Hereafter we will refer to such energies as the RNA Mfold values. It is possible to further reduce this number from 10 to 8 independent parameters by considering the circular symmetry of the NN model \cite{goldstein1992many, licinio2007irreducible}. This symmetry yields additional self-consistent relations for the dimer occupancies along the duplex: out of the 10 NNBP energies 2 can be expressed as linear combinations of the remaining 8 \cite{gray1970new,licinio2007irreducible,huguet2017derivation}. The circular symmetry property has been tested and the 8 parameters derived in DNA unzipping experiments \cite{huguet2010single,huguet2017derivation}.

\begin{figure*}[ht]
\centering
\includegraphics[width=.85\textwidth]{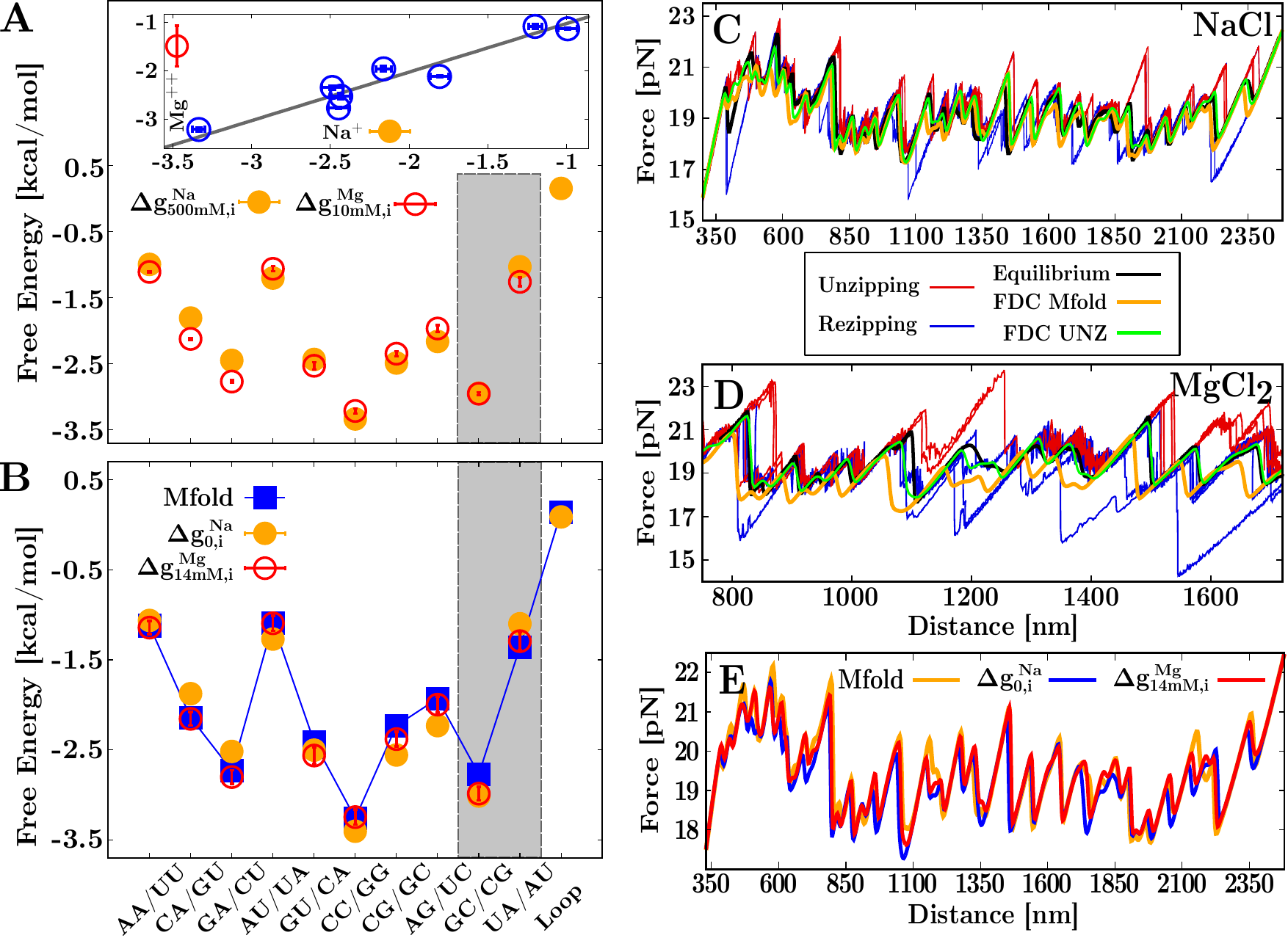}
\caption{\label{fig:results} NNBP free-energy parameters. (\textbf{A}) \textbf{Main.} Measured energies at 500mM NaCl (orange) and 10mM $\rm MgCl_2$ (red). \textbf{Inset.} Plot of the energies in $\rm Mg^{++}$ against those in $\rm Na^{+}$. The fit according to \eqref{eq:SaltFit} (gray line) gives the coefficient $a=77\pm49$ (see text). (\textbf{B}) Comparison of the Mfold energies (blue) with the 1M NaCl and (the equivalent) 14mM $\rm MgCl_2$ free-energy sets. The two parameters resulting from considering the circular symmetry have been highlighted (gray band). The loop free-energy in magnesium has not been measured (see text). Note that in sodium the error is smaller than the size of the symbol. (\textbf{C}, \textbf{D}) Comparison of the unzipping, rezipping and equilibrium FDCs (in red, blue and black, respectively) measured in 500mM NaCl and 10mM $\rm MgCl_2$ with the theoretical prediction obtained from Mfold (orange), and the energies reported in columns 1 and 2 of Table \ref{tab:NNBPenergies} (green). Mfold agrees better for sodium than for magnesium. (\textbf{E}) Comparison between the theoretical FDCs computed at the equivalent salt conditions $\Delta g^{\rm Na}_{0,i}$ (orange), $\Delta g^{\rm Mg}_{14 \rm mM,i}$ (red) and Mfold (blue) (columns 3,5,6 of Table \ref{tab:NNBPenergies}).}
\end{figure*}

We derived the eight RNA NNBP and loop energies from the equilibrium FDCs in sodium and magnesium by using a Monte Carlo optimization algorithm, analogously to the DNA case \cite{huguet2010single,huguet2017derivation}. The elastic parameters of the model include the persistence and contour-lengths of the hybrid DNA/RNA handles ($P_{\rm DNA/RNA}=10$nm and $L_{\rm DNA/RNA}=7.8$nm) and those of the ssRNA ($P=0.805$nm and interphosphate distance $l_d=0.68$nm).
The results, averaged over the different molecules, are summarized in Table \ref{tab:NNBPenergies} (columns 1,2) and plotted in Fig.\ref{fig:results}A (Main). The last two NNBP values (GC/CG and UA/AU) are obtained by applying the circular symmetry. These values support the validity of a salt equivalence rule between sodium and magnesium. To derive the rule we plotted the measured energies in [Mg$^{++}$] = 0.01 M as a function of the energies in [Na$^+$] = 0.5 M fitting them to the relation
\begin{equation}
\label{eq:SaltFit}
\Delta g^{\rm Mg}_i([\text{Mg}^{++}]) = \Delta g^{\rm Na}_i([\text{Na}^+]) - m \cdot \log{\left( \frac{[\text{Na}^+]_{\text{eq}}}{[\text{Na}^+]}\right)} \, ,
\end{equation}
where $[\text{Na}^+]_{\text{eq}}\equiv a \times [\text{Mg}^{++}]$ is the magnesium concentration in sodium equivalents and $a$ is the equivalence factor. $\Delta g^{\rm Mg}_i([\text{Mg}^{++}])$ and $\Delta g^{\rm Na}_i([\text{Na}^{+}])$ are the experimentally derived energies of motif $i$ in (Mg$^{++}$) and (Na$^{+}$) at the respective salt concentrations in molar units. Finally, $m = 0.10 \pm 0.01$ kcal/mol is the NNBP-homogeneous monovalent salt correction parameter experimentally derived in \cite{bizarro2012non},
\begin{equation}
    \Delta g_i^{\rm Na}([\text{Na}^+]) = \Delta g^{\rm Na}_{0,i} - m \cdot \log{([\text{Na}^+])} \, .
   \label{eq:SaltCorrection}  
\end{equation}
A least-squares fit to the data gives $a=77\pm49$ (Fig.\ref{fig:results}A, Inset), which is compatible with the value $a\approx100$ of previous studies \cite{bizarro2012non}. We expect that \eqref{eq:SaltFit}, with $a$ constant over a broad range of magnesium concentrations, holds if Mg$^{++}$ correlations and competitive effects between sodium and magnesium are weak. This implies diluted magnesium solutions, i.e. $[\text{Mg}^{++}]< 0.05 \rm M$ \cite{tan2007rna,tan2006nucleic}. With added sodium, Mg$^{++}$ effects dominate when $R= \rm \sqrt{[ Mg^{++}]}/[ Na^+] > 0.22 \rm M^{-1/2}$ \cite{owczarzy2008predicting}, which is the case in our experimental conditions ($R = 2 \rm M^{1/2}$). 

Given the measured energies (columns 1,2 in Table \ref{tab:NNBPenergies}), we calculated the NNBP and loop values at the reference salt conditions of 1M NaCl ($\Delta g^{\rm Na}_{0,i}$) and 1M MgCl$_2$ ($\Delta g^{\rm Mg}_{0,i}$). By combining \eqref{eq:SaltFit} and \eqref{eq:SaltCorrection}, we get
\begin{equation}
    \Delta g_i^{\rm Mg}([\text{Mg}^{++}]) = \Delta g_i^{\rm Na}([\text{Na}^+]_{\text{eq}})= \Delta g_i^{\rm Na}(a \times [\text{Mg}^{++}]) \, .
    \label{eq:NaEquiv}
\end{equation}
The resulting energies in sodium and magnesium are given in columns 3 and 4 of Table \ref{tab:NNBPenergies}, respectively. 

For a direct comparison with the Mfold set, we use \eqref{eq:NaEquiv} to report the energies at $\rm 14mM \,\, Mg^{++}\equiv \rm 1M \,\, Na^+$ (Column 5 in Table \ref{tab:NNBPenergies}), obtained by using \eqref{eq:NaEquiv}. Column 6 in the table shows the ten independent RNA Mfold energies plus the loop free-energy. The last two NNBP values (indicated in brackets) are obtained from the circular symmetry relations applied on the other eight Mfold parameters. Notice that the Mfold value for GC/CG (-3.82) is very different from our value in sodium (-3.01, column 3). This discrepancy arises from the use of eight parameters in our model while Mfold uses ten. Interestingly, by applying the circular symmetry property to the Mfold set we get for GC/CG the value -2.77, which is in better agreement with our value (-3.01). Notice that the free energy of the loop in magnesium is not given in the table as this value cannot be measured due to the inaccessibility of the last part of the unzipping curve. Results in Table \ref{tab:NNBPenergies} (columns 3,5,6) are plotted in Fig.\ref{fig:results}B which shows the overall agreement between the unzipping free-energy values and those of Mfold. For the total hybridization free-energy of the RNA hairpin the unzipping values predict $\Delta G^{\text{Na}}_0 = 4031$ kcal/mol (1M sodium) and $\Delta G^{\text{Mg}}_{14 \rm mM} = 4082$ kcal/mol (14mM of equivalent magnesium). These numbers compare well to the Mfold value $\Delta G^{\text{Mfold}} = 4086$ kcal/mol ($1\%$ relative error). The predicted FDCs computed with our free-energies (columns 1 and 2 in Table \ref{tab:NNBPenergies}) agree better with the experimental data than Mfold does, particularly for magnesium (green and orange lines versus the black line in Fig.\ref{fig:results}C,D). A comparison of the theoretical FDCs predicted by the Mfold set with those obtained with our energies at 1M NaCl and 14mM $\rm MgCl_2$ is shown in Fig.\ref{fig:results}E.

The salt rule for equivalent thermodynamics in sodium and magnesium does not necessarily imply an equivalent rule for kinetics. In Ref.\cite{bizarro2012non} a sequence known as CD4 hairpin was studied over three decades of monovalent and divalent salt concentrations in the diluted regime. Yet, the average unzipping force in magnesium was larger than in sodium at equivalent salt concentrations (Fig.S10 in the Supp. Info.).

%
\setlength{\tabcolsep}{1.5mm}
\begin{table*}[t]
\centering
\caption{\label{tab:NNBPenergies} Experimentally derived NNBP and loop RNA energies at T = 298 K}
\begin{tabular}{ccc||cccc}
& (1) & (2) & (3) & (4) & (5) & (6) \\
\textbf{NNBP} & $\rm \mathbf{\Delta g^{Na}_{500mM,i}}$ & $\rm \mathbf{\Delta g^{Mg}_{10mM,i}}$ & $\rm \mathbf{\Delta g^{Na}_{0,i}}$ & $\rm \mathbf{\Delta g^{Mg}_{0,i}}$ & $\rm \mathbf{\Delta g^{Mg}_{14mM,i}}$ & \textbf{Mfold}\\ 
\toprule
\textbf{AA/UU} & -0.99 (6) & -1.11 (1) & -1.06 (6) & -1.57 (5) & -1.14 (7) & -1.12\\
\textbf{CA/GU} & -1.81 (6) & -2.12 (1) & -1.88 (6) & -2.58 (5) & -2.15 (7) & -2.14\\
\textbf{GA/CU} & -2.45 (7) & -2.77 (2) & -2.52 (7) & -3.23 (5) & -2.80 (7) & -2.73\\
\textbf{AU/UA} & -1.20 (4) & -1.06 (4) & -1.27 (4) & -1.52 (6) & -1.09 (8) & -1.09\\
\textbf{GU/CA} & -2.43 (6) & -2.53 (6) & -2.50 (6) & -2.99 (7) & -2.56 (9) & -2.41\\
\textbf{CC/GG} & -3.33 (4) & -3.21 (4) & -3.40 (4) & -3.67 (6) & -3.25 (8) & -3.26\\
\textbf{CG/GC} & -2.45 (7) & -2.35 (4) & -2.56 (7) & -2.81 (6) & -2.38 (8) & -2.23\\
\textbf{AG/UC} & -2.16 (5) & -1.96 (5) & -2.23 (5) & -2.42 (7) & -2.00 (9) & -1.93\\
\midrule
\textbf{GC/CG} & -2.94 (8) & -2.95 (2) & -3.01 (8) & -3.41 (5) & -2.99 (8) & -3.82 [-2.77]\\
\textbf{UA/AU} & -1.03 (10) & -1.26 (7) & -1.10 (10) & -1.72 (8) & -1.29 (10) & -1.36 [-1.37]\\
\midrule
\textbf{Loop} & 0.16 (3) & --- & 0.09 (3) & --- & --- & 0.14\\
\bottomrule
\end{tabular}
\caption*{\normalfont (\textbf{Columns 1,2}) Experimentally measured NNBP energies in 500mM NaCl and 10mM $\rm MgCl_2$, respectively. The last two values (GC/CG, UA/AU) have been computed with the circular symmetry. (\textbf{Columns 3,4}) NNBP values reported at the standard conditions of 1M NaCl and 1M MgCl$_2$, respectively. (\textbf{Column 5}) NNBP energies in magnesium reported at the concentration equivalent to $\rm 1M \,\, Na^+ \equiv 14mM \,\, Mg^{++}$. (\textbf{Column 6}) Mfold prediction for the ten independent NNBP energies at 1M NaCl. NNBP values computed with circular symmetry are also reported (square brackets). Note the loop free-energy in magnesium is not given (see text). All energies are in kcal/mol and have been reported with the statistical error computed over the different molecules (in parenthesis). NNBP follow the standard notation (ex., CA/GU stands for 5$^{\prime}$-CA-3$^{\prime}$ hybridized with 5$^{\prime}$-UG-3$^{\prime}$).}
\end{table*}
%

\subsection*{Stem-Loop structures and barrier energy landscape}
\begin{figure*}[ht]
\centering
\includegraphics[width=.85\textwidth]{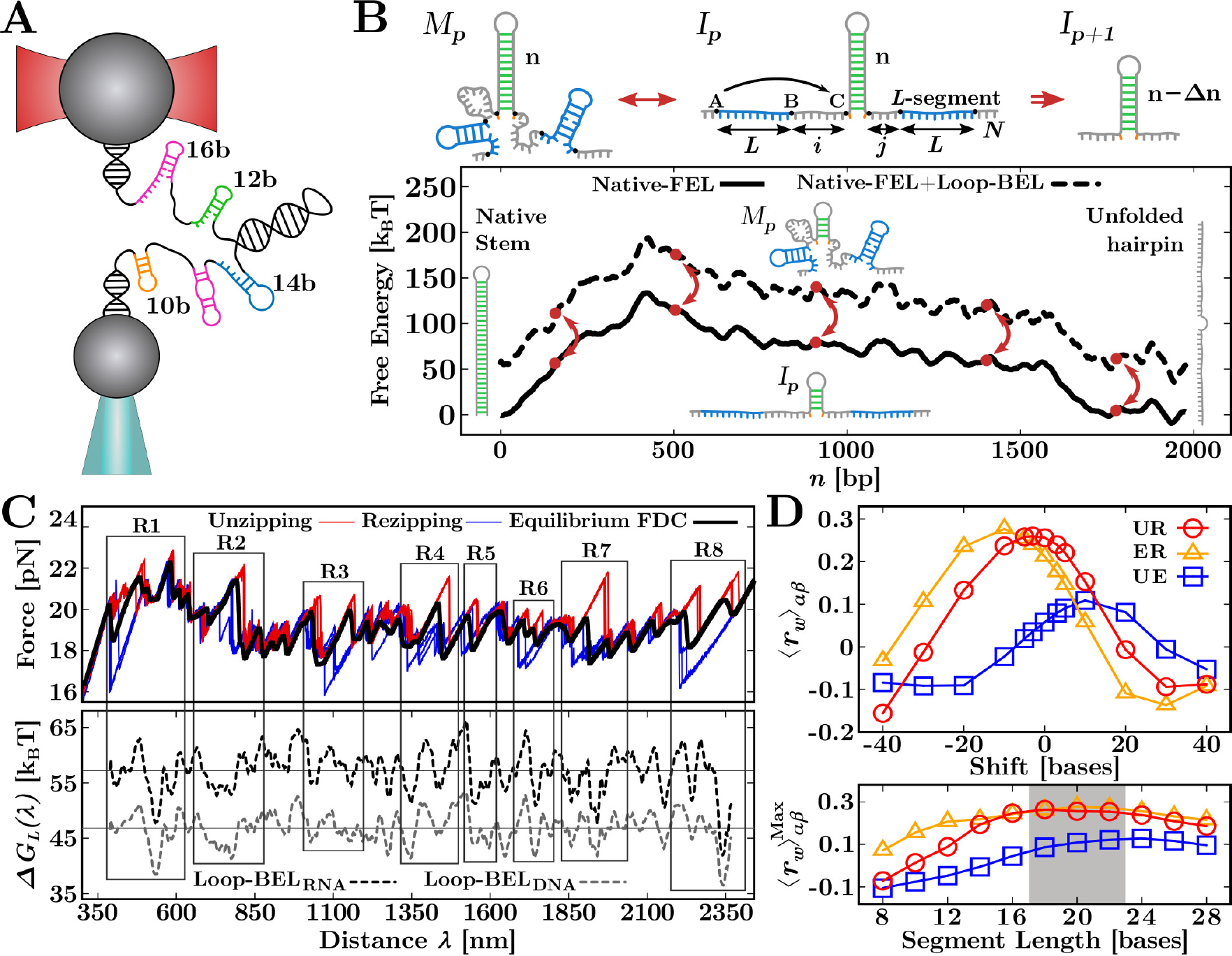}
\caption{\label{fig:hyst_loops} Stem-loops barrier energy landscape and hysteresis. \textbf{(A)} Formation of the stem-loops during the unzipping (rezipping) process. Segments of different lengths (represented with different colours) along each RNA single strand form transient stem-loop structures. 
\textbf{(B) Top.} Transition between intermediates $I_p$ and $I_{p+1}$ (double red arrow). The formation of the off-pathway (misfolded) structures consisting of stem-loops ($M_{p}$) kinetically traps (left-right red arrow) the RNA at $I_{p}$ slowing down transitions $I_{p}\to I_{p+1}$ ($I_{p}\to I_{p-1}$) during unzipping (rezipping). \textbf{Bottom.} Loop-BEL (dashed line) computed with \eqref{eq:LoopBEL} for $L=20$ added to the native-FEL of the hairpin (solid line). For a fixed $n$, the loop-BEL is the free-energy difference between structures $I_{p}$ and $M_p$ (equal to the vertical distance between red points). Red arrows depict the kinetic trapping effect induced by the loop-BEL.
\textbf{(C)} Experimental FDCs in 500mM NaCl (top) and loop-BELs at 19pN (bottom) computed for the RNA hairpin (dashed black line) and the equivalent DNA sequence (dashed gray line) for $L=20$ bases. The mean values of the loop-BEL (solid lines) are also shown. Loop-BEL minima correlate with the hysteresis regions R1-R8. 
\textbf{(D) Top. Average rolling correlation $\langle r_w \rangle_{\alpha \beta}$ as a function of the shift $s$ between loop-BEL and hysteresis profile for $\alpha \beta \equiv \rm UR, ER, UE$. \textbf{ Bottom.} Maximal average rolling correlation $\langle r_w \rangle^{\rm Max}_{\alpha \beta}$ (corresponding to $s \approx 0,-10,+10$ for UR, ER and UE, respectively) for each $L$. Hysteresis is found to be maximally correlated with stem-loops of length $L\sim [18,22]$ bases.}}
\end{figure*}
Figures \ref{fig:NaCl},\ref{fig:MgCl2} show that hysteresis is larger in magnesium than in sodium due to the longer lifetime of the intermediates in magnesium. We hypothesize a scenario where the formation of stem-loop structures along the unpaired RNA strands kinetically trap the observed intermediates $I_p$ for a given number $n$ of formed base pairs along the FDC. The size of the force jumps observed in Figs.\ref{fig:NaCl},\ref{fig:MgCl2} indicate a number $\Delta n \in [50-150]$ of unzipped-rezipped bps between consecutive intermediates.  
The large stacking free energy of RNA loops facilitates the formation of stem-loop structures at forces as high as 20pN where rezipping occurs (Fig.\ref{fig:hyst_loops}A). The stabilizing effect induced by loop formation has been demonstrated in experiments of blocking oligos in nucleic acids hairpins. By hybridizing to the complementary loop region these oligos prevent the formation of the native stem \cite{bosco2014elastic,manosas2010active}. Stem-loops often contain hairpin-like folds with non-canonical base-pairs (each colored structure in Fig.\ref{fig:hyst_loops}A corresponds to a different number of bases) stabilized by stacking and base pairing interactions. To form the native stem the two single strands pulled under opposite forces must come close to each other. However, this process facilitates the formation of off-pathway (misfolded) stem-loop structures in the single strands. In Fig.\ref{fig:hyst_loops}B (top) we depict the hairpin unzipping at position $n$ (middle) between two consecutive intermediates $I_p \rightarrow I_{p+1}$, which is slowed down by the transient formation of off-pathway (misfolded) structures ($M_p$, left) consisting of stem-loops in the single strands (blue segments) that kinetically trap the RNA. The intermediate $I_{p+1}$ (right) is rescued upon releasing $\Delta n$ bases forming the stem-loops ($M_p$) that kinetically trap the hairpin. Notice that kinetic trapping also occurs during rezipping for transitions $I_{p} \rightarrow I_{p-1}$. In the reversible regions intermediates have very short lifetimes and are not observed, meaning that kinetic trapping and hysteresis effects are negligible at the experimental pulling speeds.

The irreversibility of the unzipping-rezipping reaction can be understood by introducing a many-valley barrier energy landscape (BEL) that, for a given $n$, accounts for the off-pathway competing folds that can be formed in each single strand. We stress that the BEL is a non-standard free energy landscape describing the propensity of the hairpin to become kinetically trapped at a particular value of $n$ by off-pathway conformations of high kinetic stability. The complexity of including all possible structures is enormous, therefore we have restricted the analysis to the single stem-loops (loop-BEL) stabilized by stacking and base pairing.
Let us consider all consecutive segments of $L$ bases along each of the two unpaired RNA strands (referred to as $a$ and $b$). Let $\mathcal{S}_L^{(a,b)}$ be the set of all segments of length $L$ contained in each strand of the RNA hairpin, $\mathcal{S}_L^{(a,b)}= \{[b_i,b_{i+L}];1\leq i \leq N' = N-L\}$, where $b_i$ and $b_{i+L}$ stand for the initial and final base of the segment on strands $(a,b)$ ($N$ being the total number of bases in the hairpin). For a given $L$-segment $[b_i, b_{i+L}]$ there are several competing folds, most of them stabilized by short complementary stems plus one or more loops of varying sizes (mostly 3-8 bases). We have searched for the optimal fold of lowest free energy, $\epsilon^0_{L,i}$, by using the \textit{DINAmelt web application} \cite{markham2005dinamelt,markham2008unafold} based on Mfold. This yields the optimal set of energies $\{\epsilon^0_{L,i}\}^{(a,b)}$ for $\mathcal{S}_L^{(a,b)}$ at standard conditions. 
With the optimal set of stem-loop energies for a given $L$, we defined the loop-BEL at force $f$ and position $n$ as 
\begin{equation}
\begin{aligned}
    \label{eq:LoopBEL}
    &\Delta G_L(n,f) = \\
                    &= - k_BT \log \sum_{i,j=0}^{N-n} \exp{\left(-\frac{\Delta g^{(a)}_L(i,f) + \Delta g^{(b)}_L(j,f)}{k_BT}\right)} ,
\end{aligned}
\end{equation}
where $\Delta g^{(a,b)}_L(i,f)$ is the total free-energy contribution per strand $(a,b)$ of a stem-loop forming at distance $i$ from the junction $n$ at force $f$. Note that in \eqref{eq:LoopBEL} we assumed that all $L$-segments at the back of the junction are already hybridized into the native stem and do not contribute to the loop-BEL (green bps in Fig.\ref{fig:hyst_loops}B,top). The term $\Delta g^{(a,b)}_L(i,f)$ is given by,
\begin{equation}
\label{eq:Strand_term}
    \Delta g^{(a,b)}_L(i,f) = -\epsilon^{0(a,b)}_{L,i} + \int_0^f{x_{L+i}(f')df'} \, ,
\end{equation}
where $\epsilon^{0(a,b)}_{L,i}$ is the (positive) free energy of formation of the stem-loop at zero force along strand $(a,b)$ and the integral stands for the energy cost to bring the $L+i$ bases from A to C at force $f$ (Fig.\ref{fig:hyst_loops}B, top). The latter penalizes stem-loops that are formed far away from the junction because they cannot kinetically trap the stretched RNA. It has been computed with the Worm-Like Chain (WLC) model \cite{bustamante1991entropic} 
\begin{equation}
\label{eq:WLC}
    f_{L+i}(x) = \frac{k_BT}{4P} \left[\left(1-\frac{x}{(L+i)l_d}\right)^{-2} - 1 + \frac{4x}{(L+i)l_d} \right] ,
\end{equation}
with $l_d=0.68 \rm nm$ the interphosphate distance \cite{murphy2004probing, bizarro2012non} and $P=0.805 \rm nm$ the RNA persistence length \cite{bizarro2012non,camunas2016elastic}. To calculate the integral in \eqref{eq:Strand_term} we inverted \eqref{eq:WLC} \cite{severino2019efficient}. Note that \eqref{eq:Strand_term} equals the free energy difference between structures $M_{p}$ and $I_{p}$ in Fig.\ref{fig:hyst_loops}B, top. 

We computed $\Delta G_L(n,f)$ at the average unzipping force $f\approx 19 \rm \, pN$ at 500mM NaCl for $L$-segments in the range $L=[8,28]$, with $L=8$ the minimum number of bases needed to form stem-loops. In Fig.\ref{fig:hyst_loops}B (bottom) we show the native free-energy landscape (FEL), $\Delta G_{\rm Native}(n,f)$ (relative to the - fully unzipped - random coil state) as a black continuous line. The contribution by the loop-BEL for $L=20$ has been added to the native FEL (dashed line) to stress the fact that it kinetically traps off-pathway stem-loop structures at fixed $n$ (red arrows). The dashed line for the loop-BEL emphasizes that this is a kinetic trapping landscape that does not describe transitions between contiguous $n$ values.
In Fig.\ref{fig:hyst_loops}C (bottom) we show the loop-BEL $\Delta G_L(\lambda)$ (dashed black line) for $L=20$ together with the experimental FDC (top). The position $n$ along the sequence in \eqref{eq:LoopBEL} has been converted to trap-pipette distance $\lambda$ by using the elastic parameters, $\Delta G_L(\lambda)\equiv \Delta G_L(n,f)$. The position of the loop-BEL minima shows a correlation with the FDC regions of largest hysteresis (indicated by rectangles $R1-R8$). To compare with the DNA case, we computed the loop-BEL for the DNA analogous 2027bp sequence (obtained by replacing uracils by thymines) at the predicted average unzipping force ($\sim 16.4 \rm pN$) at 500mM NaCl \cite{huguet2010single}. Despite the profiles appear to be similar, the average barrier energy in DNA ($\sim 47\rm k_B T$, solid gray line) is lower than in RNA ($\sim 57\rm k_B T$, solid black line) because of the lower DNA unzipping force (which yields a lower elastic contribution in \eqref{eq:Strand_term}). We stress that the loop-BEL is overestimated as we have considered a restricted set (single stem-loops) among all possible competing structures. The lower the loop-BEL, the more stable the competing structures and the larger the irreversibility effects. The larger hysteresis in RNA apparently correlates with the higher kinetic stability of the stem-loops for RNA.

\subsection*{Correlation of hysteresis with stem-loops formation}
To quantify the correlation between the loop-BEL and the hysteresis, we introduced the hysteresis profile at position $\lambda$ as a measure of the dissipated work over a given distance $\Delta\lambda$ (=3nm),
\begin{equation}
\label{eq:hyst_BEL}
    \Delta G^{\rm Hyst}_{\alpha\beta}(\lambda) = - \int_{\lambda - \frac{\Delta \lambda}{2}}^{\lambda+\frac{\Delta \lambda}{2}}  |f_{\alpha}(\lambda') - f_{\beta}(\lambda')|\, d \lambda' \, ,
\end{equation}
where $\alpha,\beta$ denote the unfolding (U), refolding (R) and equilibrium (E) FDCs, leading to three distinct profiles $\Delta G^{\rm Hyst}_{\alpha\beta}(\lambda)$ with $\alpha\beta={\rm UR,UE,ER}$. The minus sign in \eqref{eq:hyst_BEL} has been introduced to positively correlate loop-BEL minima (maxima) with maximal (minimal) hysteresis. \eqref{eq:hyst_BEL} has been averaged over several cycles and different molecules. Given the loop-BEL, $\Delta G_L(\lambda) \equiv \Delta G_L(n,f=19\rm pN)$ in \eqref{eq:LoopBEL}, and the hysteresis profile, $\Delta G_{\alpha}^{\rm Hyst}(\lambda)$, we computed the Pearson correlation coefficient $r_w(\lambda) \in [-1,1]$ over a given spatial window of size $w$ as a function of $\lambda$. $r_w(\lambda) = 1$ $(r_w(\lambda) = -1)$ indicates fully correlated (anticorrelated) landscapes in that region. Correlation profiles $r_w(\lambda)$ have been calculated for $\Delta G_{\alpha \beta}^{\rm Hyst}(\lambda)$ with $\alpha \beta \equiv \rm UR,UE,ER$ (Fig.S4 in Supp. Info.). To assess the correlation between the loop-BEL and the hysteresis profile $\alpha \beta$, we defined the average rolling correlation, $\langle r_w \rangle_{\alpha \beta}$, as the average taken over the entire landscape $r_w(\lambda)$. Another parameter for the correlation analysis is $\phi_{\alpha \beta}$, defined as the probability that $r_w(\lambda) \geq 0.5$ at a given $\lambda$ averaged over the entire landscape.  Although this parameter is a better estimator of positive correlations (see Fig.S5 in Supp. Info.), here we show the standard average rolling correlation, $\langle r_w \rangle_{\alpha \beta}$. We used a sliding window of size $w \approx 100$nm, the result being insensitive to $w$ as far as it is comparable to the typical number of bases released in a force rip along the FDC ($\sim$ 50-150 bases) \cite{huguet2009statistical} (Fig.S6 in the Supp. Info.). In Fig.\ref{fig:hyst_loops}D (top) we show $\langle r_w \rangle_{\alpha \beta}$ as a function of the shift $s$ (in bases) of the loop-BEL relative to the hysteresis profiles. $\langle r_w \rangle_{\alpha \beta}$ has been calculated for the L-segment length $L=20$ at which correlation is maximal (see below). A positive shift $s>0$ means that we are testing the correlation with the loop-BEL in the rezipped region close to the junction (green bp in Fig.\ref{fig:hyst_loops}B, top), whereas a negative shift $s<0$ implies testing the correlation with the loop-BEL ahead of the junction in the unzipped region (grey and blue bp in Fig.\ref{fig:hyst_loops}B, top). Remarkably, maximum correlation is found for $\alpha \beta \equiv \rm UR$ and $s=0$ (red circles in Fig.\ref{fig:hyst_loops}D, top) showing that stem-loops formation and hysteresis are highly correlated precisely at the junction. The position of the maximum in $\langle r_w \rangle_{\alpha \beta}$ shifts to $s>0$ $(s<0)$ for $\alpha\beta \equiv \rm UE$ (ER) (blue squares and orange triangles respectively, Fig.\ref{fig:hyst_loops}D, top). We notice that for $\alpha \beta \equiv \rm ER$ the maximum in $\langle r_w \rangle_{\rm ER}$ is shifted leftwards by $s \approx -10$ bases (orange triangles) and its value almost coincides with the $\alpha \beta \equiv \rm UR$ case ($\langle r_w \rangle^{\rm Max}_{\rm ER} \sim \langle r_w \rangle^{\rm Max}_{\rm UR} \approx 0.25$, red circles). Therefore, the formation of stem-loops at a distance of $\sim$ 10 bases in the unzipped region slows down the refolding of the hairpin leading to the hysteresis observed during the rezipping process. In contrast, the maximum of $\langle r_w \rangle_{\rm UE}$ (blue squares) is shifted rightwards ($s \approx +10$) with $\langle r_w \rangle^{\rm Max}_{\rm UE}\approx 0.1<\langle r_w \rangle^{\rm Max}_{\rm UR}\approx 0.25$ (red circles). The asymmetry between UE and ER demonstrates that the largest source of irreversibility in the unzipping-rezipping experiment is the refolding process. Analogously, the rightwards shift ($\sim +10$ bases) in $\langle r_w \rangle^{\rm Max}_{\rm UE}$ is related to breathing of stem-loops and the hysteresis effects observed in the unfolding FDCs. Finally, we analyzed the dependence of $\langle r_w \rangle^{\rm Max}_{\alpha \beta}$ with the length $L$ of the segments forming the stem-loops (Fig.\ref{fig:hyst_loops}D, bottom). All curves show a broad maximum for $L \approx 18-22$, meaning that this is the characteristic size of the stem-loops size that kinetically trap the RNA intermediates during unzipping and rezipping.

These results are supported by various control analyses. In Fig.S7 of the Supp. Info. we report the average rolling correlation between loop-BEL and hysteresis for different hairpins obtained by shuffling segments of the original sequence and with random sequences. This comparison shows a positive correlation for the original hairpin sequence, $\langle r_w \rangle^{\rm Max}_{\rm UE} \simeq 0.25$, whereas for the shuffled and random control sequences correlations are apparently lower (in the range [-0.06:+0.07] and [-0.04:-0.005] for the two controls in Fig.S7A and B, respectively). Finally, we computed the Pearson coefficient between loop-BEL and hysteresis profile in the irreversible and reversible regions, separately. This analysis (Fig.S8 of the Supp. Info.) shows a positive correlation only in the irreversible regions whereas in the reversible ones correlations are spurious due to thermal fluctuations and instrumental noise.
%

\section*{Discussion and conclusions}
Detailed knowledge of the energetics of hybridization of RNA is key to determine the thermodynamic stability of RNA structures, from dsRNA to tertiary RNAs, essential in many biophysical processes. We studied the kinetics of RNA hybridization by mechanically pulling a 2kbp RNA hairpin with optical tweezers. By repeatedly unzipping and rezipping the RNA we measured the sequence-dependent FDCs in sodium and magnesium. The large hysteresis observed along the FDCs demanded nonequilibrium physics methods to derive the fully reversible FDC from the irreversible pulling data. In fact, quasi-static RNA unzipping experiments are not feasible as the lifetime of the intermediates requires pulling speeds that are exceedingly low. Estimates based on the Bell-Evans model range from 0.1nm/s to 1pm/s for irreversible hairpin segments of 30-40 bp.   

By using an optimization algorithm, we derived the free energies of the ten nearest-neighbor base pairs (NNBP) in RNA (Fig.\ref{fig:results}), finding good agreement with the Mfold values reported for sodium. To the best of our knowledge, NNBP energies are not currently available for RNA in magnesium, making our results the first available dataset. The highest difference between our energies and Mfold is found for CG/GC in sodium (Fig.\ref{fig:results}B), a relevant motif prone to methylation that accumulates in many regulatory regions \cite{cross1995cpg,esteller2008epigenetics}. Moreover, the results for magnesium show the validity of a general salt equivalence rule 80/1 for which 10mM $\rm Mg^{++}$ corresponds to 800mM $\rm Na^+$ (Fig.\ref{fig:results}A). Although the scope of this result has been tested in a single salt condition, its validity should span the dilute salt regime where cooperative salt effects are negligible ($[\rm Mg^{++}]<0.05 M$) and competition effects with sodium are weak ($R=\sqrt{[\rm Mg^{++}]}/[\rm Na^+] > 0.22 \, M^{-1/2}$). A salt equivalence rule has been disputed on the basis of experimental data obtained in bulk experiments using atomic emission spectroscopy in buffer equilibrated samples \cite{lipfert2014understanding}. Although this technique is capable of determining the fraction of cations that are dissociated and bound to the RNA, it does not provide a direct measurement of free energies. Here we have demonstrated the validity of a 80/1 salt equivalence rule at the level of individual NNBP motifs. To date, this is the most direct confirmation of the validity of the 100/1 rule of thumb for the equivalence of the non-specific binding energy of sodium and magnesium in RNA structures.

The strong hysteresis observed between the RNA unzipping and rezipping FDCs is driven by the collective effects of multiple stem-loop structures that kinetically trap the RNA. The effect is stronger in magnesium than in sodium, probably because the two charges of magnesium transiently stabilize nucleotide contacts to a higher extent. Irreversible regions are characterised by a high frequency of purine stacks and Watson-Crick bonds along the unpaired strands which lead to the multiple peaks observed in the experimental FDCs, even for forces as high as 20pN. Note that stacking alone could not transiently stabilize stem-loops at such high forces, it is necessary the concurrent formation of base pairs within each of the RNA strands. It is quite reasonable that such stem-loop structures also exhibit some degree of cooperativity, the more they proliferate the more they facilitate the formation of additional nearby stem-loops inhibiting native folding. Cooperative folding effects have been also found in DNA \cite{schneider2019sequence,viader2021cooperativity}, RNA \cite{greenleaf2008direct,halma2019complex,woodson2010compact} and proteins \cite{ritchie2015probing,shank2010folding,schug2010protein}. The intermediates $I_p$ in the unzipping-rezipping experiments are reminiscent of the cooperative foldons hypothesized to drive protein folding \cite{englander2014nature,englander2017case}. This cannot be otherwise, as the only way to form the native stem is to sequentially form the intermediates, one after the other, starting from the unfolded state. The remarkable effect of force is to increase the lifetime of the intermediates that would be difficult to detect in melting experiments. 

We have shown that the hysteresis correlates with the transient stabilization of RNA stem-loop structures along each unpaired single strand. The formation of stem-loops in the proximity of the hybridization junction stabilizes the intermediates that enhance the hysteresis observed in the FDC. By defining a stem-loops barrier energy landscape (loop-BEL, Fig.\ref{fig:hyst_loops}B), we found a correlation between the sequence regions where stem-loops are maximally stable (minima of the loop-BEL) with those where hysteresis along the FDC is large (Fig.\ref{fig:hyst_loops}C). To support this interpretation we measured the correlation between the loop-BEL and the hysteresis profiles (\eqref{eq:hyst_BEL}) as a function of the relative shift between them (Fig.\ref{fig:hyst_loops}D). We have found that the hysteresis observed in the FDCs maximally correlates with the stem-loop formation at the hybridization junction. Additional test controls on shuffled and random sequences support the statistical significance of the measured correlation. Typical stem-loop sizes of about 20 bases are responsible for the observed hysteresis effects. Interestingly, this number is similar to that of foldon residues in protein folding \cite{englander2014nature}.  
We stress that the loop-BEL as a function of $n$ is not a standard free-energy landscape as neither the trap-pipette distance $\lambda$ nor $n$ are true reaction coordinates for the stem-loops. For a given $n$ ($\lambda$) the loop-BEL is a kinetic trapping landscape that quantifies off-pathway (misfolded) configurations $M_p$ that compete with the folding intermediates $I_p$. Future work should lead to a better understanding of the stabilizing kinetics of these structures and the energy landscape describing transitions between them. We notice that along the reversible regions the signal-to-noise ratio is very low due to instrumental drift and noise effects, which are detrimental in evaluating the correlation between sequence and hysteresis.

It is remarkable that hysteresis is observed in some specific regions of the FDC but not in others. To explain this, we have searched for specific sequence motifs that promote stacking, hybridisation and stem-loop formation within each single strand. We have searched for segments of length $N\geq 6$ bases containing consecutive purines for stacking (A,G) and complementary bases for hybridization (A,U and G,C) within each single strand for the irreversible and reversible regions (see Supp. Info. for a detailed discussion). We find a higher frequency of purine stacks and hybridizing bases in the irreversible regions showing that these regions enhance stem-loops formation and hysteresis.

Finally, we have designed a short RNA hairpin of 20bp that ends in an A-rich dodecaloop to enhance stacking effects. The hairpin also contains many contiguous A,U’s along the sequence promoting base pairing within the ssRNA. If pulled under equivalent salt conditions (100nm/s, 1M NaCl and 10mM MgCl$_2$, at 298K) the native hairpin unzips around 21pN. Interestingly, in magnesium the hairpin also forms an alternative misfolded structure ($\approx 30\%$ of the times) that is seldom observed in sodium (Fig.S11 in Supp. Info.). This result demonstrates that the presence of stacking and base pairing along the ssRNA facilitates misfolding. This effect is enhanced in magnesium, showing that kinetic effects between sodium and magnesium are nonequivalent. The same experiment but with a stem that does not contain contiguous bases capable of base pairing does not show the misfolded state neither in sodium nor in magnesium (Fig.S12 in Supp. Info.). Overall, these results demonstrate that concurrent stacking and hybridisation among bases within the single strands leads to the observed irreversible effects.

Fluctuation relations have proven to be a fabulous playground to extract equilibrium information from irreversible pulling experiments in molecular structures from native RNAs \cite{collin2005verification,hummer2010free} to proteins \cite{li2008rna} and ligand binding \cite{zhuang2005single}. Moreover, the well-defined reaction coordinate of the unzipping process shows that intermediates stabilization is induced by the formation of stem-loops along the RNA single strands. These results suggest that stem-loops formation is an essential step in RNA folding in \textit{in vitro} and \textit{in vivo} conditions. Indeed, numerical and theoretical studies of RNA folding models have emphasized the importance of loop formation in the hybridization reaction \cite{hyeon2007mechanical,einert2008impact,einert2011theory,einert2011secondary}. This might contribute to explain a wide range of RNA behaviors, from misfolding \cite{alemany2012experimental} and multiplicity of native structures \cite{gralla1974biological}, to the RNA thermostatic and cold-denaturation phenomenon \cite{mikulecky2004heat,iannelli2020cold}. Ultimately, the promiscuity of transiently stable RNA structures might be related to the diversity of physiological responses observed when such RNAs interact with the human genome, as in the case of the RNA viruses. 
%
%
\matmethods{
\subsection*{Molecular synthesis}
We synthesized an RNA hairpin made of a stem of 2027 equally represented canonical Watson-Crick base pairs (the occurrence of each NNBP motif is reported in Table S2 of the Supp. Info.), ending in a tetraloop and inserted between short hybrid DNA/RNA handles (29bp). Short handles ensure a sufficiently large signal-to-noise force and a slower unzipping/rezipping kinetics \cite{forns2011improving} facilitating the detection of the intermediates occurring along the FDC. 
RNA constructs of a few kbp in length with a specific sequence are difficult to synthesize. In fact, the attempts to synthesize the hairpin as a single transcript from plasmids containing two copies of a DNA fragment coding for the hairpin stem failed probably due to hairpin nuclease SbcCD proteins recognizing long palindromic sequences and introducing double-stranded breaks (DSB) on them \cite{eykelenboom2008sbccd}.
To circumvent this problem, we devised a synthesis protocol by which two RNA molecules, RNA1 and RNA2, are synthesized separately and then covalently joined using T4RNA ligase 2 (Fig.S1 in Supp. Info.). RNA1 contains a small $5^{\prime}$-sequence (region 1.1) that pairs with a digoxigenin-labeled DNA oligonucleotide to form a small DNA/RNA handle, a larger portion (region 1.2) which anneals with a reverse complementary strand from RNA2 molecule (region 2.2) to form the hairpin stem region, and a $3^{\prime}$-sequence (region 1.3) that contains the GAAA tetraloop. Apart from region 2.2, RNA2 molecule also contains a $3^{\prime}$-sequence used to form a second small DNA/RNA handle after annealing with a biotinylated DNA oligonucleotide. A detailed description of the synthesis is reported in the Supp. Info.
\subsection*{Recovery of the equilibrium FDC}
The experimental FDCs show strong irreversibility localized in 8 regions in sodium and 4 regions in magnesium (Figs.\ref{fig:NaCl} and \ref{fig:MgCl2}, respectively). Each region is 
limited by starting (left, $L$) and ending (right, $R$) trap positions where the RNA is in equilibrium and exhibits intermediates $I_p$ being $p=(1,2,\ldots, P)$. Let $\mathcal{S}$ be the set of $P+2$ states containing $L, R$ and the intermediates, $\mathcal{S} = (I_0 = L,I_1,I_2\dots,I_P,I_{P+1}=R)$. In our setting, during the forward process (F) the trap position $\lambda$ is moved at a constant speed starting in $I_0$ at $\lambda_0$ and ending in $I_p$ at $\lambda$. Similarly, in the time-reverse protocol (R) the trap position is moved back at the same velocity starting in $I_p$ at $\lambda$ and ending in $I_0$ at $\lambda_0$. Thus, the extended fluctuation relation reads \cite{junier2009recovery,alemany2012experimental}
\begin{equation}
\label{eq:EFR}
\frac{\phi^{I_0 \rightarrow I_p}_F}{\phi^{I_p \rightarrow I_0}_R} \frac{P^{I_0 \rightarrow I_p}_F(W)}{P^{I_p \rightarrow I_0}_R(-W)} = \exp{\left[ \frac{W-\Delta G_{I_0 I_p}(\lambda)}{k_B T} \right]} \, ,
\end{equation}
where $P^{I_0 \rightarrow I_p}_F(W)$ $(P^{I_p \rightarrow I_0}_R(-W))$ is the partial distribution of the work $W$ (defined as the work distribution conditioned to states $I_0,I_p$) measured along the F (R) protocol, $\Delta G_{I_0 I_p}(\lambda) = G_{I_p}(\lambda) - G_{I_0}(\lambda_0)$ is the free-energy difference between states $I_p$ at $\lambda$ and $I_0$ at $\lambda_0$ and $\phi^{I_0 \rightarrow I_p}_F$ $(\phi^{I_p \rightarrow I_0}_R)$ is the fraction of paths along F(R) starting in $I_0$ ($I_p$) at $\lambda_0$ ($\lambda$) and ending in $I_p$ ($I_0$) at $\lambda$ ($\lambda_0$). $k_B$ is the Boltzmann constant and $T$ is the temperature. Note that all trajectories in the reverse process end up in $I_0$ whatever is the initial state $I_p$ so that $\phi^{I_p \rightarrow I_0}_R = 1$. For a finite number of trajectories, a direct extrapolation of $\Delta G_{I_0 I_p}(\lambda)$ from \eqref{eq:EFR} leads to biased results. Therefore, we developed a method based on the combination of the extended Bennett acceptance ratio method \cite{junier2009recovery} and the (extended) Jarzynski estimator \cite{jarzynski1997nonequilibrium} to extract the best estimate for $\Delta G_{I_0 I_p}(\lambda)$ (see Supp. Info. for the detailed description).

Given the energies of all the states occurring in a region, the equilibrium free-energy is recovered as the potential of mean force taken over all the free-energy branches so that
\begin{equation}
\label{eq:eqEnergy}
\Delta G_{\rm eq}(\lambda) = -k_B T \log \left( \sum_{p=0}^{P+1} \exp{-\bigl (\frac{\Delta G_{I_0 \rightarrow I_p}(\lambda)}{k_BT}\bigr)} \right) \, .
\end{equation}
Eventually, the equilibrium FDC of the irreversible region is computed as $f_{\rm eq}(\lambda) = \partial \Delta G_{\rm eq}(\lambda)/\partial \lambda$. The recovery of the equilibrium FDC in all the irreversible regions in sodium and magnesium allowed us to reconstruct the whole equilibrium FDCs.}
\showmatmethods{} 

\acknow{P.R. was supported by the \textit{Angelo della Riccia} foundation. C.V.B. is Research Career Awardee of the National Research Council of Brazil (CNPq) and was partly financed by the Coordenação de Aperfeiçoamento de Pessoal de Nível Superior, Brasil (CAPES), Finance Code 001. F.R. was supported by Spanish Research Council Grants FIS2016-80458-P, PID2019-111148GB-I00 and the Institució Catalana de Recerca i Estudis Avançats (ICREA) Academia Prizes 2013 and 2018.}
\showacknow{} 


\bibliography{BibliographyMAIN}

\end{document}


\maketitle

\SItext

\subsection*{Hairpin synthesis}

\begin{figure*}
\centering
\includegraphics[width=.9\textwidth]{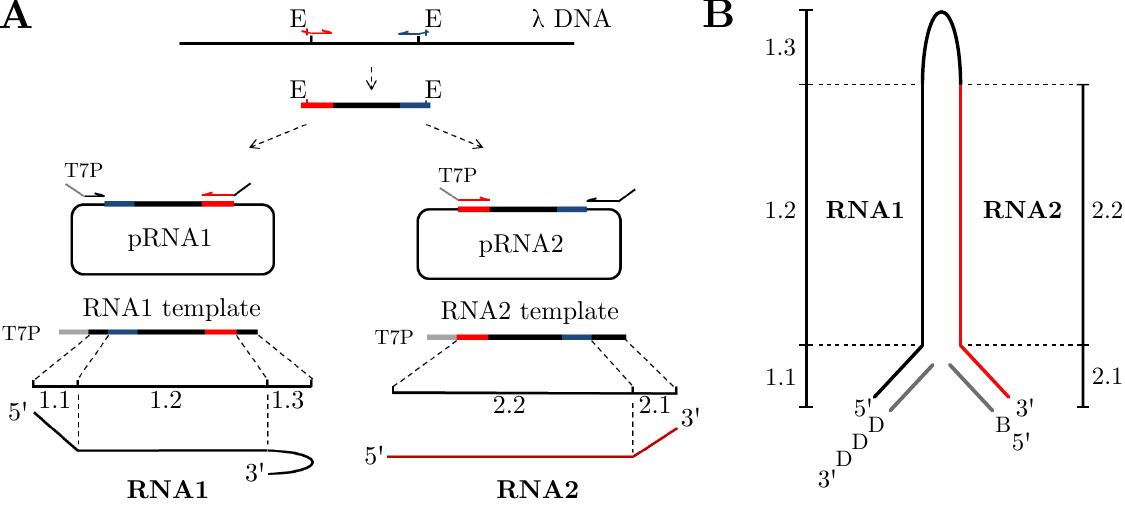}
\caption{\label{fig:synthesis}\textbf{(A)} RNA hairpin synthesis. A PCR amplicon of 2027bp in length obtained from $\lambda$-DNA and containing EcoRI sites (E) at its termini was cloned into pBR322 plasmid in both orientations, generating the pRNA1 and pRNA2 constructs. The pRNA1 and pRNA2 constructs were then used as templates for PCR reactions. Both PCR products contained a minimal T7 RNA Polymerase Promoter (T7P) carried by the forward primers. The PCR products were used as templates for \textit{in vitro} transcription reactions that were performed to synthesize RNA1 and RNA2 molecules. Regions 1.1 and 2.1 are derived from pBR322 sequence, regions 1.2 and 2.2 from $\lambda$-DNA sequence, and region 1.3 from the RNA1 reverse primer. \textbf{(B)} RNA hairpin structure and assembly. The hairpin is formed by annealing molecules RNA1 and RNA2, a digoxigenin (DIG)-labeled and a biotin (BIO)-labeled oligonucleotide. RNA1 molecule contains three regions: region 1.1 anneals with DIG-labeled oligonucleotide to form RNA1 handle. Region 1.2 anneals with RNA2 and together with region 2.2 from RNA2 forms most of the hairpin stem. Finally, region 1.3 forms the hairpin loop and the upper part of the stem. Apart from region 2.2, RNA2 molecule also contains a $3^{\prime}$ portion (region 2.1) that anneals with BIO-labeled oligonucleotide to form the RNA2 handle.}
\end{figure*}

We synthesized an RNA hairpin made of a stem of 2027 equally represented canonical Watson-Crick base pairs (Table \ref{tab:BaseFreq}), ending in a GAAA tetraloop and inserted between short hybrid DNA/RNA handles (29bp). The synthesis of long RNA hairpins is a challenging task (see main text) and required the development of a suitable protocol which can be split into 7 main steps and is schematically depicted in Fig.\ref{fig:synthesis}A,B.

\subsubsection*{PCR amplification of target sequence}
The target sequence was selected inside a $\lambda$-DNA (J02459) region (30286 – 38650) that was previously shown to be efficiently transcribed in both strands by T7 RNA polymerase, the same enzyme used in our experiments \cite{abels2005single}. A PCR reaction was performed to obtain an amplicon of 2027bp in length using $1 \mu \text{M}$ of Univ\texttt{\_}hairpin\texttt{\_}F and 1$\mu$M of EcoRI\texttt{\_}2.0kb\texttt{\_}R primers (Table \ref{tab:oligos}), 25ng of $\lambda$-DNA (Dam-) as DNA template, 1.5mM MgCl$_2$, 1X Opti and 1X HiSpec buffers, dNTPs 0.2mM each, and 4U of Eco Taq Plus DNA Polymerase (Ecogen). Primer sequences were selected using Primer3Plus software \cite{untergasser2007primer3plus}. EcoRI sites were added at $5^{\prime}$ termini of both primers (Table \ref{tab:oligos}, sequence in bold). Cycling parameters were as follow: initial denaturation step (94$^{\circ}$C) for 1 min 30 sec, enzyme addition (hotstart), 30 cycles of denaturation at 94$^{\circ}$C for 45 sec, annealing at 60$^{\circ}$C for 1 min and extension at 72$^{\circ}$C for 6 min, with a final extension step at 72$^{\circ}$C for 7 min.

\subsubsection*{Synthesis of pRNA1 and pRNA2 constructs}

The 2kbp PCR amplicon was digested with EcoRI (NEB, New England Biolabs) and cloned into vector pBR322/EcoRI \cite{bolivar1977constructionI, bolivar1977constructionII}. Plasmid DNA was purified using Illustra PlasmidPrep Mini Spin Kit (GE Healthcare). The insert orientation was evaluated by digesting plasmids with EcoRV and according to it constructs were defined as pRNA1 or pRNA2 (Fig.\ref{fig:synthesis}). These constructs were used as templates for PCR reactions using primers RNA1\texttt{\_}T7Forw and RNA1\texttt{\_}Rev (pRNA1) or RNA2\texttt{\_}T7Forw and RNA2\texttt{\_}Rev (pRNA2) (Table \ref{tab:oligos}).

\subsubsection*{PCR amplification of templates for \textit{in vitro} transcription}

The pRNA1 and pRNA2 constructs were used as templates for PCR reactions using primers RNA1\texttt{\_}T7Forw and RNA1\texttt{\_}Rev (pRNA1) or RNA2\texttt{\_}T7Forw and RNA2\texttt{\_}Rev (pRNA2) (see Table \ref{tab:oligos}). RNA\texttt{\_}T7Forw primer contains a cytidilate nucleotide (in bold) upstream the minimal T7 RNA Polymerase Promoter ($5^{\prime}$-ctaatacgactcactataggga-$3^{\prime}$) to improve transcription efficiency \cite{baklanov1996effect}, followed by a pBR322-annealing sequence ($5^{\prime}$-ataaaaataggcgtatcacgag-$3^{\prime}$). This sequence codes for part of RNA1 handle. Primer RNA1\texttt{\_}Rev anneals at its $3^{\prime}$ termini ($5^{\prime}$-gaaaaacgcctcgagtgaag-$3^{\prime}$) with the Univ\texttt{\_}hairpin\texttt{\_}F binding site located at the end of pRNA1 insert opposite to RNA1\texttt{\_}T7Forw binding site (see Fig.\ref{fig:synthesis}). The $5^{\prime}$ termini of RNA1\texttt{\_}Rev ($5^{\prime}$-ctcatctgtttccagatgag-$3^{\prime}$) codes for the last 8bp of the RNA hairpin near the loop and the GAAA tetraloop itself (in bold, reverse complement). The sequence $5^{\prime}$-ggga-$3^{\prime}$ was introduced between $5^{\prime}$ and $3^{\prime}$ portions of RNA1\texttt{\_}Rev in order to introduce the sequence $5^{\prime}$-uccc-$3^{\prime}$ into RNA1. This tetranucleotide RNA sequence is located between the hairpin stem portion formed by sequences from the 2kbp insert and the last 8bp-stem and loop regions coded by RNA1\texttt{\_}Rev sequence, and serve to base pair the first four nucleotides ($5^{\prime}$-ggga-$3^{\prime}$) at the $5^{\prime}$ end of RNA2 molecule.   
RNA2\texttt{\_}T7Forw anneals with same sequence that pairs with RNA1\texttt{\_}Rev primer ($5^{\prime}$-gaaaaacgcctcgagtgaag-$3^{\prime}$), but it is used in PCR reactions with pRNA2 construct. As in the case of RNA1\texttt{\_}T7Forw, its $5^{\prime}$ termini contain an optimized T7 promoter containing an upstream “c” nucleotide ($5^{\prime}$-ctaatacgactcactataggga-$3^{\prime}$). RNA2\texttt{\_}Rev primer contains an pBR322-annealing region and codes for the RNA2 handle. PCR reactions were performed using the same conditions previously described. Amplification products were purified from PCR mixtures using the GFX PCR DNA and Gel Band Purification Kit (GE Healthcare).

\subsubsection*{\textit{In vitro} transcription of RNA1 and RNA2 molecules}

\textit{in vitro} transcription reactions were performed using the T7 MEGAscript-High Yield transcription Kit (ThermoFisher Scientific/Ambion) according to manufacturer's recommendations. Samples were incubated with 3$\mu$L of TURBO DNase (2U/$\mu$L) for 15 min at 37$^{\circ}$C, and synthesized RNA was precipitated by addition of 90$\mu$L of LiCl Precipitation solution (7.5M lithium chloride, 50mM EDTA). Reactions were incubated overnight at -20$^{\circ}$C, centrifuged for 15 min at 13,000 rpm, washed twice with 70$\%$ ethanol and resuspended in 15$\mu$L of nuclease-free water.

\subsubsection*{Treatment of RNA2 molecules with Calf Intestinal Phosphatase (CIAP) and Polynucleotide Kinase (PNK)}

\textit{in vitro} transcribed RNA 2 molecules were treated with 1U of CIAP (Roche) for 1h at 50$^{\circ}$C to remove their $5^{\prime}$ triphosphate ends. Dephosphorylated RNA2 molecules were treated with Polynucleotide Kinase (PNK) to produce RNA molecules containing $5^{\prime}$ monophosphate termini according to manufacturer's recommendations. The reactions were heat-inactivated by incubating for 20 min at 65$^{\circ}$C and precipitated with LiCl as described in the following section.

\subsubsection*{Digoxigenin $3^{\prime}$ tailing of S Handle A oligonucleotide}
\label{sec:tailing_S_handle}

S Handle A (Table \ref{tab:oligos}) tailing with digoxigenins were performed by using the DIG oligonucleotide Tailing Kit 2nd Generation (Roche), according to manufacturer's recommendations. DIG-labeled S Handle A was purified by using the Qiaquick Nucleotide Removal kit (Qiagen).

\subsubsection*{Assembling RNA1 and RNA2 molecules to form the 2kbp RNA hairpin}

The assembly of RNA hairpin was performed in one annealing step, where RNA1, CIAP and PNK-treated RNA2, S Handle A and biotin-labeled S Handle B2 oligonucleotides (Table \ref{tab:oligos}) were incubated together. A total of 20$\mu$g of RNA1 and 20$\mu$g of CIAP/PNK-treated RNA2 were incubated with 5$\mu$L of DIG-tailed S Handle A (2$\mu$M), 5$\mu$L of $5^{\prime}$-Bio-S Handle B2 (2$\mu$M), 2$\mu$L of Tris 1M, pH 7.0, 2$\mu$L of NaCl 5M and water to a final volume of 80$\mu$L. Reactions were incubated for 1 h at 65$^{\circ}$C and cooled to 10$^{\circ}$C at a rate of 0.5$^{\circ}$C/min using a thermocycler. After a final cooling step at 10$^{\circ}$C for 1 h and 30 min, the samples were subjected to microdialysis. The annealing reaction was pipetted over a 0.05$\mu$m Millipore membrane which was put in a plate containing 50mL of 20mM Tris.HCl, 5mM NaCl, pH 7.5, and allowed to stay for 1 h. Microdialyzed, annealed molecules were then incubated with 1$\mu$L of T4 RNA ligase 2 (RNL2 1U/$\mu$L) (NEB) and 1X RNL2 Reaction Buffer for 2 h at 37$^{\circ}$C to covalently join RNA1 and RNA2 molecules. Ligated RNA hairpin molecules were microdialyzed against Tris.HCL 100mM, EDTA 1mM as described above and stored at -20$^{\circ}$C or directly used in single-molecule experiments.

\section*{Recovery of the equilibrium FDC}

To recover the equilibrium FDC from the irreversible experimental data, we developed an approach based on the extended fluctuation relations (EFR). Here we report in detail the implementation of this method. 

Let us first recall the notation and the relations introduced in the main text. Given an irreversible region of the FDC limited by starting (left, $L$) and ending (right, $R$) equilibrated states, let $\mathcal{S}$ be the set of all the states in that region $\mathcal{S} = (I_0=L,I_1,I_2\dots,I_{P},I_{P+1}=R)$ being $(I_1,\dots,I_P)$ the partially equilibrated intermediates. During the experimental forward process (F) the trap position $\lambda$ is moved at a constant speed starting in $I_0$ at $\lambda_0$ and ending in $I_p$ at $\lambda$. Similarly, in the time-reversed protocol (R) the trap position is moved back at the same speed starting in $I_p$ at $\lambda$ and ending in $I_0$ at $\lambda_0$. Thus, the EFR reads
\begin{equation}
\label{eq:EFR}
\frac{\phi^{I_0 \rightarrow I_p}_F}{\phi^{I_p \rightarrow I_0}_R} \frac{P^{I_0 \rightarrow I_p}_F(W)}{P^{I_p \rightarrow I_0}_R(-W)} = \exp{\left[ \frac{W-\Delta G_{I_0 I_p}(\lambda)}{k_B T} \right]} \, ,
\end{equation}
where $P^{I_0 \rightarrow I_p}_F(W)$ $(P^{I_p \rightarrow I_0}_R(-W))$ is the partial distribution of the work $W$ measured along the F (R) protocol, $\Delta G_{I_0 I_p}(\lambda) = G_{I_p}(\lambda) - G_{I_0}(\lambda_0)$ is the free-energy difference between states $I_p$ at $\lambda$ and $I_0$ at $\lambda_0$ and $\phi^{I_0 \rightarrow I_p}_F$ $(\phi^{I_p \rightarrow I_0}_R)$ is the fraction of paths starting in $I_0$ ($I_p$) at $\lambda_0$ ($\lambda$) and ending in $I_p$ ($I_0$) at $\lambda$ ($\lambda_0$). $k_B$ is the Boltzmann constant and $T$ is the environment temperature.

\begin{figure*}
\centering
\includegraphics{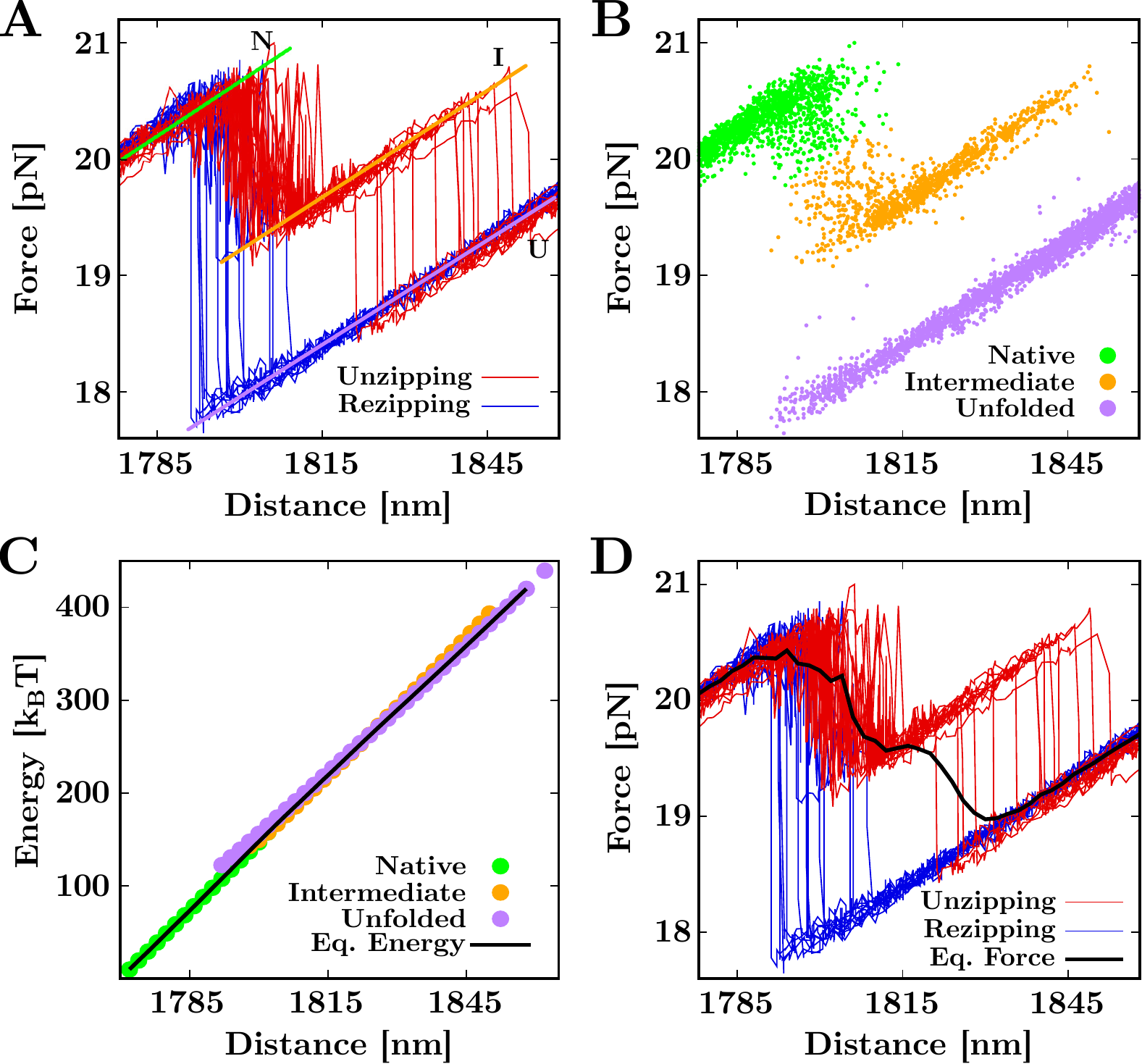}
\caption{\label{fig:reconstruction_steps} Reconstruction of the equilibrium FDC in the 3-states region measured in 500mM NaCl (see Fig.3 in the main text). The protocol involves four main steps. \textbf{(A)} Given the unzipping/rezipping experimental trajectories, each state (Native, Intermediate, Unfolded) is identified and labeled (green, orange and purple lines, respectively). \textbf{(B)} The experimental points are assigned to the closest state through a least-square fit. \textbf{(C)} The free energy of each state (U,I,N) is computed as the result of the combination of the Bennett and Jarzynski equations (see text). The equilibrium free-energy between all the states (black line) is computed through \eqref{eq:eqEnergy}. \textbf{(D)} Eventually, the equilibrium FDC (black line) is recovered by computing \eqref{eq:eqFDC}.}
\end{figure*}

Let us now introduce the extended Bennett acceptance ratio (EBAR) method \cite{bennett1976efficient,shirts2003equilibrium}. By multiplying \eqref{eq:EFR} by the function $f(W) = \left(1 + \frac{\phi_R}{\phi_F} \frac{n_F}{n_R} \exp{\frac{W-\Delta G_{I_0 I_p}}{k_BT}} \right)^{-1}$ and integrating over the work one gets
\begin{equation}
\label{eq:EBARmethod}
\frac{u}{k_B T} = - \log{\left( \frac{\phi^{I_0 \rightarrow I_p}_F}{\phi^{I_p \rightarrow I_0}_R} \right)} + z_R(u) - z_F(u) \, ,
\end{equation}
where the variance of the free-energy estimator is minimized by the equations
\begin{subequations}
\label{eq:EBAReqs}
\begin{align}
z_R(u) &=  \log{\left \langle \frac{\exp{-\frac{W^R_i}{k_BT}}}{1 + \frac{\phi_R}{\phi_F} \frac{n_F}{n_R} \exp{-\frac{W^R_i+u}{k_BT}}} \right \rangle_{R}} = \log{ \frac{1}{n_R} \sum_{i=1}^{n_R} \left( \frac{\exp{-\frac{W^R_i}{k_BT}}}{1 + \frac{\phi_R}{\phi_F} \frac{n_F}{n_R} \exp{-\frac{W^R_i+u}{k_BT}}} \right)}\\
z_F(u) &= \log{ \left \langle \frac{1}{1 + \frac{\phi_R}{\phi_F} \frac{n_F}{n_R} \exp{\frac{W^F_i-u}{k_BT}}} \right \rangle}_{F} = \log{ \frac{1}{n_F} \sum_{i=1}^{n_F} \left( \frac{1}{1 + \frac{\phi_R}{\phi_F} \frac{n_F}{n_R} \exp{\frac{W^F_i-u}{k_BT}}} \right)}
\end{align}
\end{subequations}
for the choice $u=\Delta G_{I_0 I_p}$. Here $\langle \cdot \rangle_{F(R)}$ denotes the thermodynamic average over the forward (reverse) process and $n_{F(R)}$ is the number of forward (reverse) trajectories that are in state $p$ for a given $\lambda$. Note that $n_{F(R)}$ corresponds to the total number of trajectories in a region (roughly a hundred per region, as reported in the main text) only at equilibrium ($I_0$), whereas, for a given $\lambda$, it ranges from 1 to 20-25 per state. Although these numbers are small, it has been shown that they lead to reasonable free energy estimates when applying the extended fluctuation relations to analogous cases \cite{alemany2012experimental,camunas2017experimental}. Thus, \eqref{eq:EBARmethod} is a self-consistent relation which returns the free energy of the transition $I_0 \rightarrow I_p$ at position $\lambda$ by using information from both the folding and refolding trajectories (bidirectional estimator). Importantly, if, for given $\lambda$, a state $I_p$ only occurs in the forward (reverse) process, $n_R=0$ ($n_F=0$) in \eqref{eq:EBAReqs} causing \eqref{eq:EBARmethod} to fail. Because hysteresis differently affects the forward and the reverse processes, the number of intermediates is, in general, different between unzipping and rezipping, making EBAR unsuitable to our purpose. This is clearly shown in Fig.\ref{fig:reconstruction_steps}A, where the unzipping FDCs always exhibit three states whereas the majority of the rezipping trajectories directly go from $R$ to $L$.

\begin{figure*}
\centering
\includegraphics{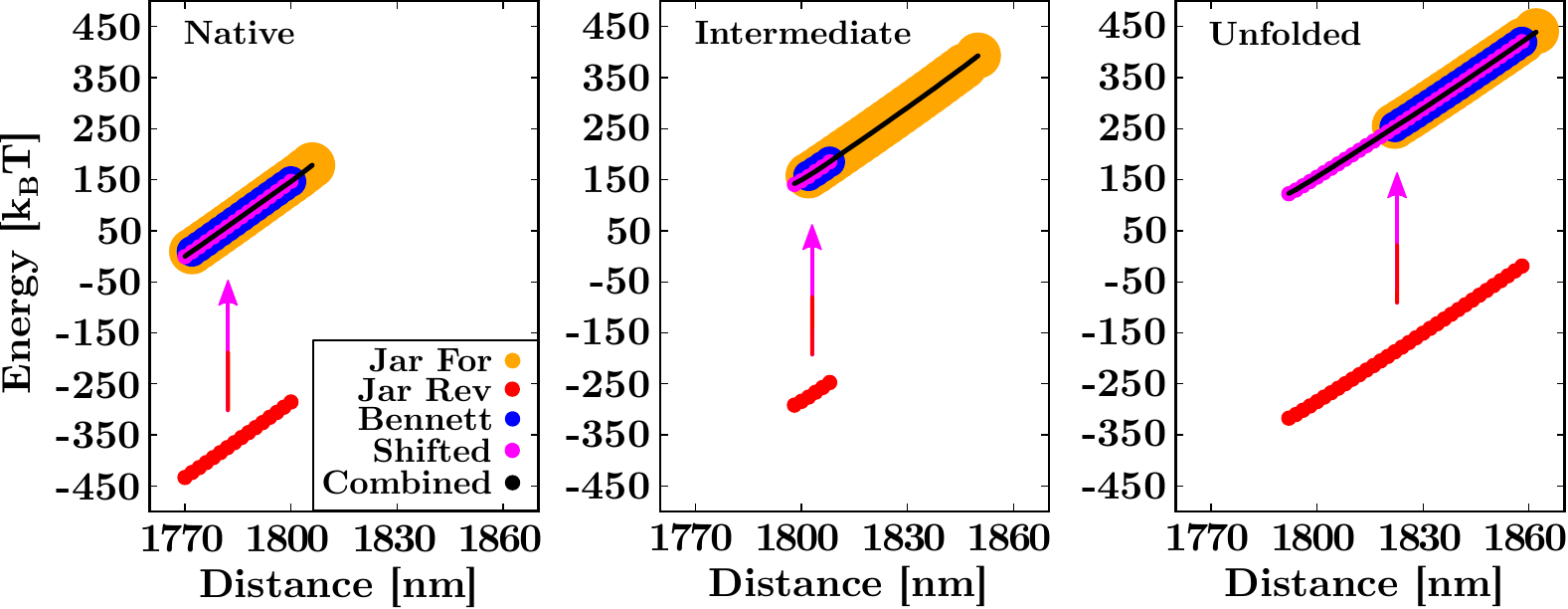}
\caption{\label{fig:Jarz&Bennet} free energy computation of the 3-states (Native, Intermediate, Unfolded) region showed in Fig.\ref{fig:reconstruction_steps}. The free energy of each state is computed by combining \eqref{eq:EBARmethod} and \eqref{eq:EJExt}. EBAR method (blue dots) only holds if $n_F,n_R \neq 0$ for each position $\lambda$, often leading to free energy estimations limited to a restricted data fraction (see the Intermediate state panel). The forward and reverse Jarzynski estimators are used to compute the energies of the forward (yellow dots) and reverse (red dots) trajectories for each $\lambda$. The (biased - see text - ) results are eventually corrected (violet dots) according to the computed EBAR values, used as reference. This procedure gives the complete free-energy set of each state (black dots).}
\end{figure*}

To solve this problem we introduced the extended form of the Jarzynski free-energy estimator \cite{jarzynski1997nonequilibrium} 
\begin{equation}
\label{eq:EJExt}
\left \langle \exp{ -\frac{W_{F(R)}}{k_BT}} \right \rangle _{F(R)} =  \phi^{S_{0(p)} \rightarrow S_{p(0)}}_{F(R)} \exp{ -\frac{\Delta G_{I_0 I_p}(\lambda)}{k_BT}} \, ,
\end{equation}
which allows to compute $\Delta G_{I_0 I_p}(\lambda)$ by only taking into account the trajectories of the forward (reverse) protocol that visit state $I_p$ (unidirectional estimator). 
However, unidirectional free-energy estimators have a slow convergence and need a big number of trajectories to obtain reliable free-energy measures, i.e. are affected by non-negligible bias \cite{palassini2011improving}. On the contrary, bidirectional free-energy estimators, such as the EBAR, have a faster convergence and lead to much smaller errors \cite{ytreberg2006comparison}. Therefore, we correct the bias of the free energies obtained with the Jarzynski estimator by using the results of the EBAR as reference. This has been obtained with following method. Let us consider a state $I_p$ stretching over the positions set $\{\lambda\}$. Firstly, we computed the energies $\{ \Delta G(\lambda)_{I_0 I_p}\}^{\rm EBAR}$ by solving \eqref{eq:EBARmethod} for all those $\lambda$ having $n_F(\lambda),n_R(\lambda) \neq 0$. Then, the energies of the forward, $\{ \Delta G_{I_0 I_p}(\lambda) \}^{\rm Jar}_S$, and reverse, $\{ \Delta G_{I_p I_0}(\lambda) \}^{\rm Jar}_R$, processes have been separately computed for all the $\lambda$ by using \eqref{eq:EJExt}. By using the energy values computed with the EBAR $\{ \Delta G_{I_p I_0}(\lambda) \}^{\rm EBAR}$ as reference, the closest intersection point with the Jarzynski energy sets $\{ \Delta G_{I_p I_0}(\lambda) \}^{\rm Jar}_F$ ($\{ \Delta G_{I_p I_0}(\lambda) \}^{\rm Jar}_R$) is computed. Eventually, the bias is corrected by applying a rigid shift to $\{ \Delta G_{I_p I_0}(\lambda) \}^{\rm Jar}_F$ ($\{ \Delta G_{I_p I_0}(\lambda) \}^{\rm Jar}_R$) in order to match the Bennett set. This method is schematically shown in Fig.\ref{fig:Jarz&Bennet} for each state (Native, Intermediate, Unfolded) exhibited by the 3-states region in Fig.\ref{fig:reconstruction_steps}.

Given the set $\{\Delta G_{I_0 I_p}(\lambda)\}$ for all the $I_p$, the equilibrium free energy is recovered as
\begin{equation}
\label{eq:eqEnergy}
\Delta G_{\rm eq}(\lambda) = -k_B T \log \left( \sum_{p=0}^{P+1} \exp{-\frac{\Delta G_{I_0 I_p}(\lambda)}{k_BT}} \right) \,
\end{equation}
for any $\lambda$ (black line in  Fig.\ref{fig:reconstruction_steps}C). Eventually, by solving the equation
\begin{equation}
\label{eq:eqFDC}
f_{\rm eq}(\lambda) = \frac{\partial}{\partial \lambda}\Delta G_{\rm eq}(\lambda) \, ,
\end{equation}
we computed the equilibrium FDC (black line in Fig.\ref{fig:reconstruction_steps}D). 

This method allowed us to compute the equilibrium FDC in all the 8 irreversible regions in sodium and the 4 irreversible regions in magnesium (Fig.2,3 of the main text). Eventually, the full equilibrium FDC has been recovered by the piecewise merging of the equilibrated FDC segments.

\section*{Conversion from number of bases to trap-pipette distance}

The conversion between the number of unzipped basepairs $n$ and the trap-pipette distance $\lambda$ is standard in experimental setups such as optical tweezers and AFM, where the position of the force device (optical trap and cantilever) is controlled \cite{severino2019efficient}. In a nutshell, the relative distance between the optical trap and the bead in the pipette at a given force $f$ is given by,
\begin{equation}
\label{eq:Conversion}
\lambda(f) = x_b(f) + x_h(f) + x_{\rm ssRNA}(n,f) \, ,
\end{equation}
where $x_b(f)$ is the bead displacement, $x_h(f)$ is the handle’s extension and $x_{\rm ssRNA}(n,f)$ is the ssRNA extension given by Eq.(6) (main text) with $L+i = 2n$ (equal to the total number of unzipped bases). Here only the relative value of $\lambda$ matters, however in our study we choose $\lambda=0$ for the case when the two beads of the experimental setup (bead in trap, bead in pipette) are in contact. The elastic contributions $x_b(f)$ and $x_h(f)$ are given by, $x_b(f)=f/k_b$ for a trap of stiffness equal to $k_b$ and $x_h(f)$ is given by the worm-like chain (WLC) model for the hybrid DNA/RNA handles of persistence length ($P_{\rm DNA/RNA}$ = 10nm) and contour length per base-pair ($L_{\rm DNA/RNA} = 7.8\rm nm$) that we used in the thermodynamic analysis part of the main text (Section \textit{Derivation of the NNBP energies for RNA}). The above relation between $\lambda$ and $n$ is univoque and has been used to convert $n$ into $\lambda$.

\section*{Control tests for the correlation analysis}

To corroborate the validity of the correlation analysis we performed tests on different controls. First, starting from the original RNA sequence, for each $L$-segment (in the range $L = [8,28]$) we generated a new hairpin by randomly shuffling segments of a given length $L$ along the sequence. In this way, a different shuffled sequence is obtained for each value of $L$. Given the loop-BEL of each shuffled sequence, we computed the maximum average rolling correlation $\langle r_w \rangle^{\rm Max}_{\rm UR}$ (shift $s=0$) with respect to the hysteresis between unfolding and refolding (Eq.(7) with $\alpha \beta = \rm UR$) for a window of size $w = 100$. Fig.\ref{fig:ControlTest}A shows the results of this analysis (orange circles) along with the analogous results obtained for the original sequence (blue squares). As discussed in the main text (see Fig.5D), the correlation of the original sequence increases with the stem-loops size $L$ and has a maximum at $\langle r_w\rangle^{\rm Max}_{\rm UR} \simeq 0.25$ for $L \approx 18-22$ bases. On the contrary, the correlation for the shuffled sequences varies in the range [-0.06:0.07] with no apparent trend.
As a second control, we generated three different random RNA hairpins with the same GC content of the original sequence. Then, for the case $L = 20$ (which roughly corresponds to the maximum correlation observed) we computed the loop-BEL and the average rolling correlation with the hysteresis. The results are shown in Fig.\ref{fig:ControlTest}B. Analogously to the previous case, the random sequences (red triangles) do not appear to be correlated with the hysteresis. In fact, the correlation falls in the range [-0.04:-0.005] so that the average of $\langle r_w\rangle^{\rm Max}_{\rm UR}$ over the random sequences is roughly equal to $-0.02$ which magnitude (in absolute number) is 10 times lower than the value of the original sequence ($\simeq 0.25$).

Let us point out that the measured values of the average rolling correlation (or analogous direct quantities) are, in general, much smaller than the fraction of highly correlated points (Fig.\ref{fig:HighCorrFraction}). The normalized correlation (as defined by the Pearson coefficient) between the hysteresis and the loop-BEL is sensitive to several factors. First, the hysteresis landscape computed in Eq. (7) (see main text) is much more accurate in the regions where the irreversibility is large, i.e. where the difference between the unzipping and rezipping FDCs is large. In fact, when the hysteresis is small, i.e. along the reversible regions, Brownian fluctuations and instrumental effects contribute to reducing the signal-to-noise ratio of the measured correlation of the hysteresis landscape with the loop-BEL. In the reversible regions, thermal (Brownian) fluctuations in the unzipping and the rezipping FDCs mask the correlation between the (low) hysteresis and the loop-BEL. The loop-BEL is noiseless by construction, whereas the computed hysteresis is not, so the correlation along the reversible regions is dominated by the noise. Moreover, instrumental effects are also detrimental in estimating such a correlation in the reversible regions. In fact, slight misalignments between the experimental trajectories give contributions to the measured hysteresis that are comparable with those due to residual irreversibility effects. Finally, correlation measurements require matching the experimental measure of the hysteresis profile and the loop-BEL, further reducing correlation estimates. These sources of error render the Pearson coefficient in the reversible regions inaccurate. Therefore, it is not surprising that the Pearson coefficient measured over the entire profile and the rolling one averaged over a given window are both small, whereas the one averaged over the entire profile but restricted to the irreversible regions is markedly larger as shown in Fig.\ref{fig:RegionsPearsonCorr}.

\section*{Sequence dependency of the hysteresis}

Here we show results for all segments of length $N\geq 6$ bases containing consecutive purine (for stacking) and consecutive Watson-Crick complementary bases (for hybridization) along the two unpaired strands of the RNA hairpin. The aim is to identify differences in purine content (for stacking) and Watson-Crick base pairs (for hybridization) between the two kinds of regions, to demonstrate that irreversible effects are sequence dependent. For the stacking motifs we do not discriminate between purines G and A being both counted in the same set. For example, irreversible region 1 contains 220 base pairs and a total of 3 segments of 6 consecutive purines (a single GGGGGG and GGAAAG on one strand and AGGGGA on the other strand) and 2 segments of 7 consecutive purines (AGGAGAA, AGAGAAA on one strand). An analogous count is made on segments capable of forming Watson-Crick complementary bases on the two strands by counting the number of segments containing consecutive A,U or G,C on each of the two strands, again without discriminating their specific order. For example, the same irreversible region 1 contains 1 segment of 9 consecutive G,C (CGCGGGGGG) and 1 segment of 10 consecutive G,C (CGCCGCCGCG). In contrast, the reversible region 3/4 (meaning that it separates irreversible regions 3 and 4) contains 97 base pairs and has only 1 segment of 6 consecutive A,U (UUAAAA). Table \ref{tab:SequenceAnalysis} summarises all results and Fig.\ref{fig:SequenceAnalysis} illustrates them with frequency histograms. Note that the fraction of the different segments of purine-stacks and Watson-Crick bases in Fig.\ref{fig:SequenceAnalysis} are normalized by the total number of  bases $\Delta n$ contained in each region (irreversible or reversible). Thus, for a given region, the fraction of bases of a given type (stacking or base-pairing) is defined as $f_N = (M_N N)/\Delta n$, where $M_N$ is the total number of segments of length $N$ of that type (reported in Table \ref{tab:SequenceAnalysis}). Given the values of $f_N$, for each $N$ we computed the average fraction of bases (for stacking or base-pairing) over all the irreversible (reversible) regions
\begin{equation}
\label{eq:averagef}
\overline{f_N} = \frac{1}{n_{\rm regions}}\sum_{\rm regions} f_N \, ,
\end{equation}
where $n_{\rm regions}$ is the number of irreversible (reversible) regions ($n_{\rm regions}$ = 8 and $n_{\rm regions}$, respectively). Finally, we defined the average segment length $\langle N \rangle$ of a given type (stacking or base-pairing) for the irreversible (reversible) regions as the weighted average over $\overline{f_N}$, which is
\begin{equation}
\label{eq:averageN}
\langle N \rangle = \frac{\sum_N N \overline{f_N}}{\sum_N \overline{f_N}}
\end{equation}
Analogously, the variance of the segment length for the irreversible (reversible) regions and for the two types of analysis (stacking or base-pairing) has been computed as
\begin{equation}
\label{eq:varN}
\langle N^2 \rangle - \langle N \rangle^2 = \frac{\sum_N N^2 \overline{f_N}}{\sum_N \overline{f_N}} - \left( \frac{\sum_N N \overline{f_N}}{\sum_N \overline{f_N}} \right)^2
\end{equation}
The results are shown in Fig.\ref{fig:SequenceAnalysis} and point out that stacking and base-pairing effects are larger in the irreversible regions than in the reversible ones. Overall, we found that stacking and base-pairing contribute to the observed hysteresis facilitating the formation of stem-loop structures along each single-strand.

\section*{Experiments in short RNA hairpins}

In a previous work \cite{bizarro2012non} a sequence known as CD4 hairpin was studied over three orders of magnitude of monovalent and divalent salt conditions in the diluted regime finding that the average unzipping force in magnesium is larger than the average unzipping force in sodium at the equivalent salt concentrations as derived from thermodynamics. Results are shown in Fig.\ref{fig:CD4L4} where we plot the results obtained in that reference under the light of the newly derived $77 \pm 49:1$ salt rule. As we can see the values of the folding free energies nicely match each other according to the salt rule (left panel). Also rupture forces during unzipping and rezipping fulfill the rule albeit with a small systematic difference. This is probably due to the fact that the rupture force is a nonequilibrium quantity. The salt rule, although generally satisfied for the kinetics, is not as clean as for the free energy, probably because many other factors affect kinetics as compared to thermodynamics (Fig.\ref{fig:CD4L4}, central and right panels). In particular, at equivalent monovalent salt conditions, the average unzipping force in magnesium (filled blue squares) is larger than the average unzipping force in sodium (filled orange circles). Furthermore, the average rezipping force in magnesium (empty blue squares) is larger than the average unzipping force in sodium (empty orange circles). This is observed for two different pulling rates (central panel, 1.8pN/s; right panel, 12.5 pN/s) underlining that the amount of hysteresis (related to the difference in the average unzipping and rezipping forces) is always larger for magnesium at the equivalent salt concentrations as derived from thermodynamics.

Here we compare the kinetic of the CD4 hairpin with the results obtained by pulling a short RNA hairpin sequence specifically designed to contain base stacking and hybridisation motifs, mimicking the kinetics of a single irreversible region in the RNA hairpin. The short RNA hairpin (52 bases) contains a stem of 20 complementary base pairs that ends in a dodecaloop GAAAAAAAAAAA that creates stacking between the 11 adenines. We will denote it as hairpin A (Fig.\ref{fig:HairpinA}A). This hairpin is very different from the CD4 haripin, which has fully complementary stem ending in a tetraloop (GAAA) but does not contain relevant contiguous stacking and hybridising base pair motifs along the unpaired ssRNA. In fact, unzipping experiments of the CD4 shown that only the native structure is formed. In contrast, upon pulling hairpin A under similar experimental conditions  (100nm/s, 10mM $\rm MgCl_2$ and 298K) the hairpin forms an alternative misfolded structure (red unzipping curves and green misfolded band in Fig.\ref{fig:HairpinA}B, right panel). While misfolding is commonly observed in magnesium (roughly $30\%$ frequency), in sodium it is rarely observed (red unzipping curves in Fig.\ref{fig:HairpinA}B, left panel). We stress the importance of base pairing interactions within the individual RNA strands. Interestingly, the same experiments carried out on the above mentioned CD4 hairpin but with the stem ending in a dodecaloop (that we will denote as hairpin B) does not show the misfolded state neither in sodium nor in magnesium (Fig.\ref{fig:HairpinB}). Interestingly, hairpin B has equally enhanced stacking effects as much as hairpin A does (both contain the A-rich dodecaloop), however Watson-Crick base pairing on the 7 A,U in each of the single strands at the beginning of the stem in hairpin A is not present in hairpin B. The repeated AU motif in hairpin A, plus the stacking stabilisation of the A-rich dodecaloop, both induce the formation of a competing structure. The large dispersion of unzipping forces of the misfolded state (7-17pN) might be interpreted as arising from a mechanically rigid structure with a short distance to the transition state\footnote{This result can be inferred in the Bell-Evans model in the Gaussian approximation, which shows that the standard deviation of the rupture force distribution equals $\frac{k_BT}{x^{\ddag}}\log(\frac{3+\sqrt5}{2})$ with $x^{\ddag}$ the distance to the transition state.}. Alternatively, the lower value of the average unzipping force of the misfolded state (as compared to the 22pN of the native) and its large dispersion (in the range 7-17pN) might be interpreted as due to the fact that the misfolded state is not unique. In this case, the misfolded state contains multiple competing structures stabilized by the weaker A-U bonds along the unpaired strands.



\begin{figure*}
\centering
\includegraphics{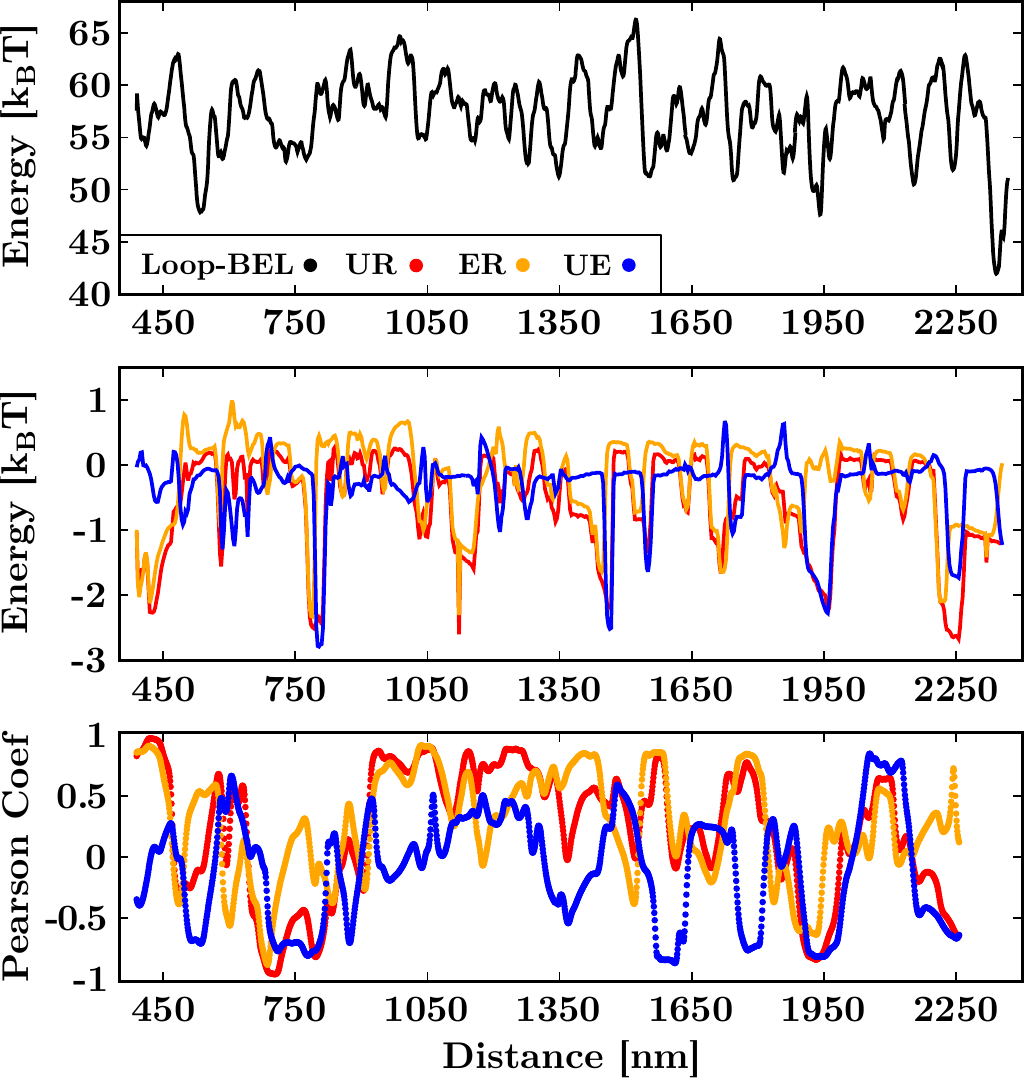}
\caption{\label{fig:PearsonTest} Correlation between the loop-BEL and the hysteresis profiles. \textbf{(Top)} Loop-BEL $\Delta G_L(\lambda)$ computed for the RNA 2kbp sequence according to Eq. (4) for $L=20$ (see main text). \textbf{(Middle)} Hysteresis profiles $\Delta G_{\alpha \beta}^{\rm Hyst}(\lambda)$ with $\alpha\beta = $ UR (red), ER (orange), UE (blue) computed for the 500mM NaCl experimental trajectories by solving Eq. (7) (see main text). \textbf{(Bottom)} Pearson correlation coefficients $r_w(\lambda)$ resulting from the comparison between loop-BEL and the $\Delta G_{\alpha \beta}^{\rm Hyst}(\lambda)$ over windows of length $w\approx100$, as described in the main text. Maximal correlation appears for $\alpha\beta = \rm UR, ER$ revealing the unzipping process as the main source of hysteresis.}
\end{figure*}

\begin{figure*}
\centering
\includegraphics{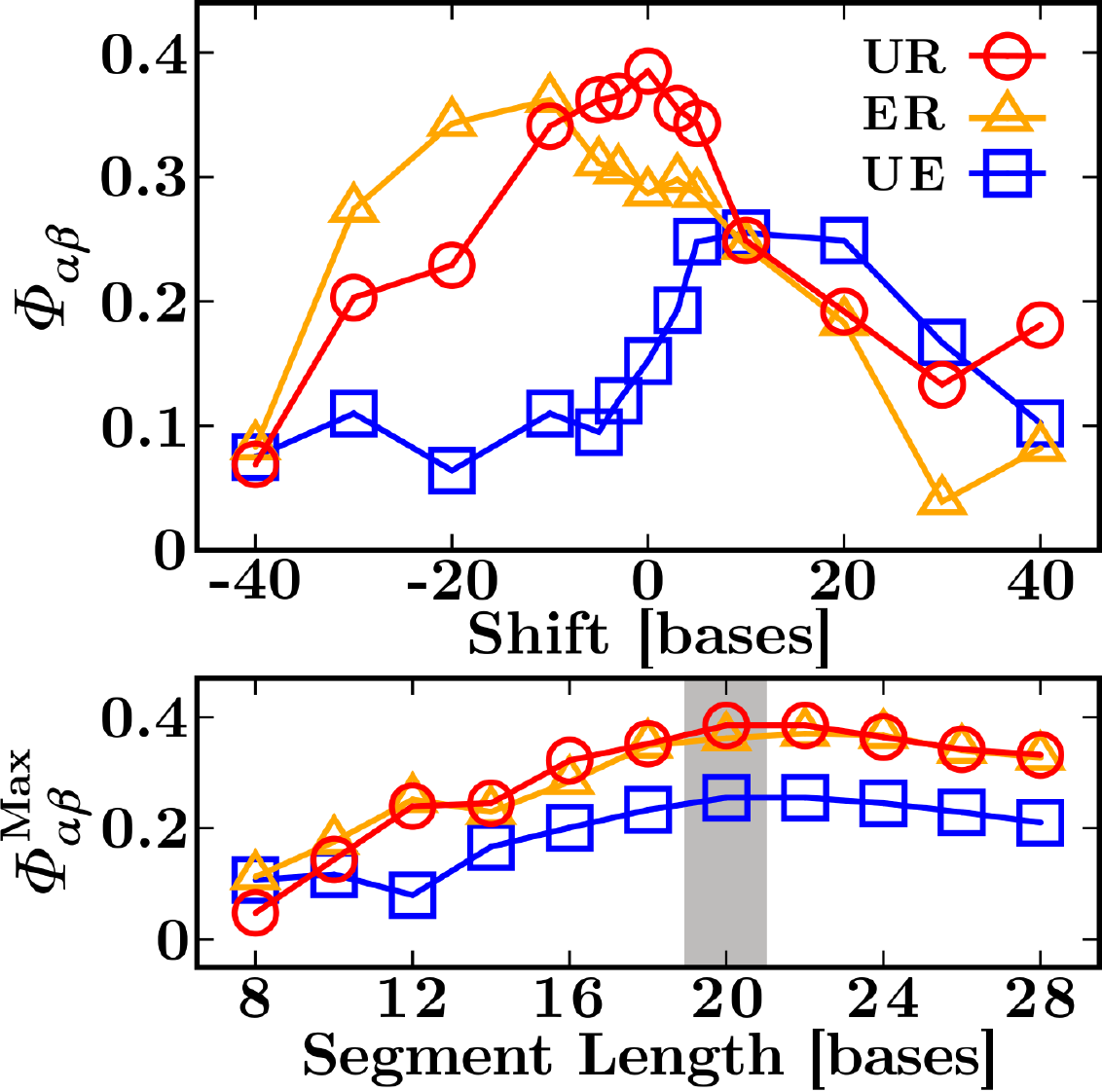}
\caption{\label{fig:HighCorrFraction} Correlation between loop-BEL and hysteresis profile evidenced by the probability $\phi_{\alpha \beta}$ that $r_w(\lambda) \geq 0.5$ at a given $\lambda$. \textbf{Top.} $\phi_{\alpha \beta}$ as a function of the shift $s$ (in bases) of the loop-BEL relative to the hysteresis profiles for the case $L=20$ (see main text). \textbf{Bottom.} Dependence of $\phi^{\rm Max}_{\alpha \beta}$ with the length $L$ of the segments forming the stem-loops. All curves exhibit a single broad maximum for $L \approx 20$, showing that this is the characteristic stem-loop size that slows down RNA folding to the native stem.}
\end{figure*}

\begin{figure*}
\centering
\includegraphics{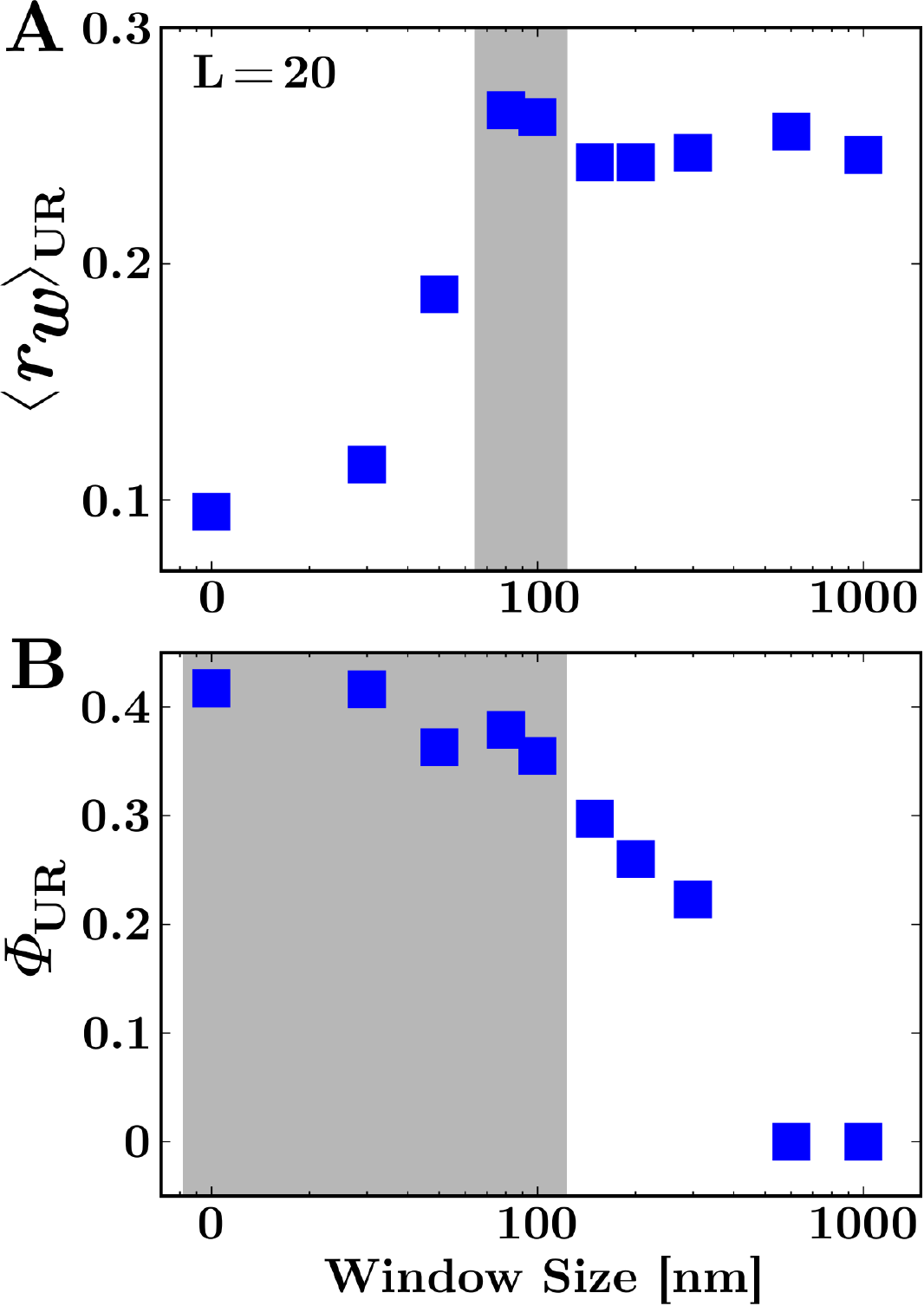}
\caption{\label{fig:WindowSize} Analysis of the optimal window size $w$ for the computation of the correlation profile $r_w(\lambda)$ in Fig.\ref{fig:PearsonTest} \textbf{(A)} Average rolling correlation $\langle r_w \rangle_{\alpha \beta}$ (computed between the loop-BEL and the hysteresis profile for $\alpha \beta = \rm UR$ and stem-loops of size $L=20$) as a function of the window size $w$. The correlation rapidly increases with $w$ and exhibits a maximum in the range [100,150]nm (grey band). \textbf{(B)} $\phi_{UR}$ as a function of the window size $w$ for the case $L=20$. The correlation is maximum in the range [10,150]nm (grey band) and is damped for larger values. Despite $\phi_{UR}$ is stable for a broader interval of $w$ than $\langle r_w \rangle_{\rm UR}$, both quantities exhibit maximal correlation at $w \sim 100$nm, which roughly correspond to the average size of the released (annealed) base-pairs during the unzipping (rezipping) process.}
\end{figure*}

\begin{figure*}
\centering
\includegraphics[width=1.\textwidth]{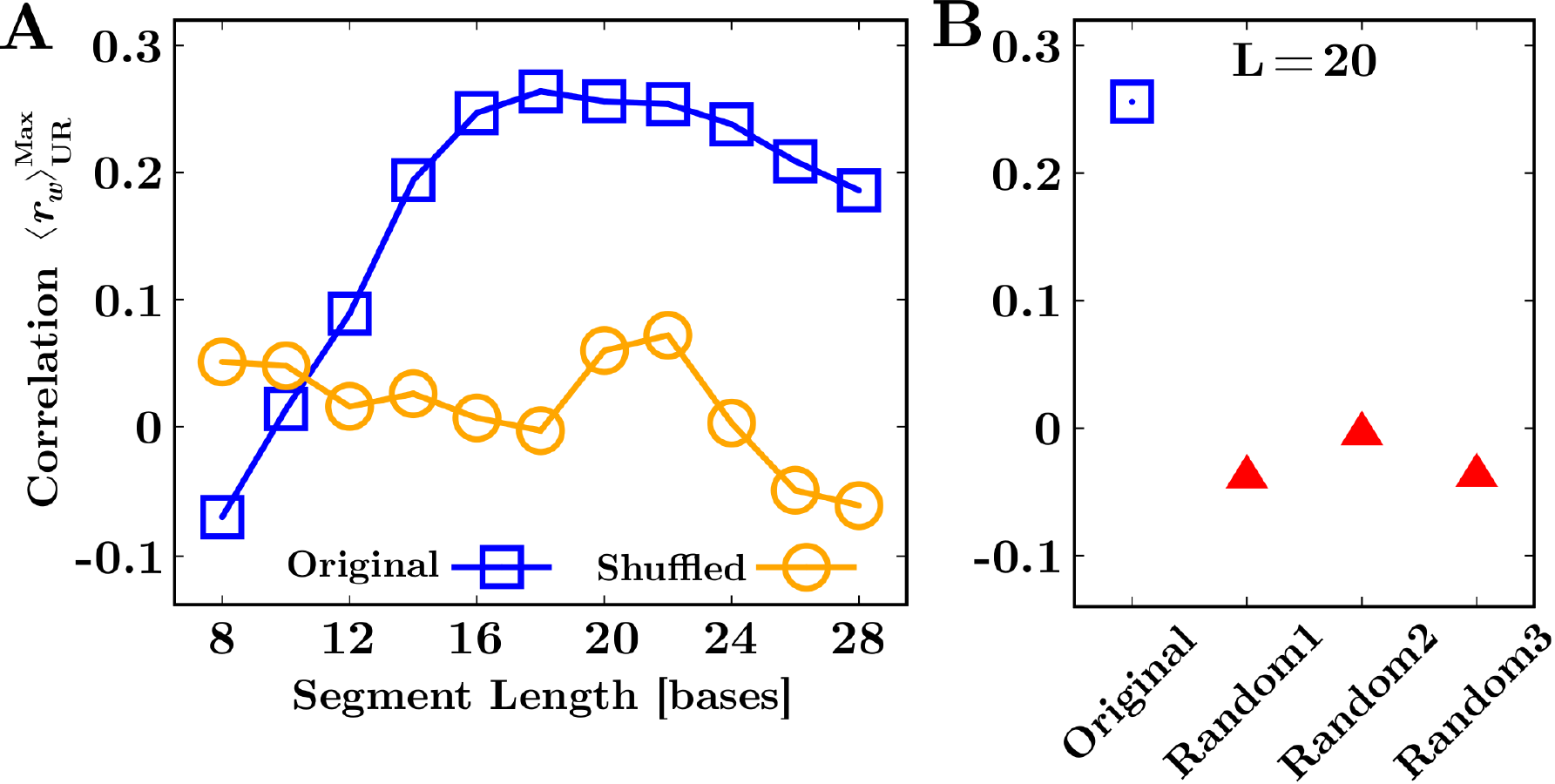}
\caption{\label{fig:ControlTest} Maximum average rolling correlation $\langle r_w \rangle^{\rm Max}_{\rm UR}$ (shift $s=0$) between hysteresis and loop-BEL for different control sequences. \textbf{(A)} Average rolling correlation computed for the shuffled sequences (orange circles) and the original sequence (blue squares) for each value of $L$=[8,28]. \textbf{(B)} Average rolling correlation of the hysteresis with the Loop-BEL for the random sequences (see text) at the maximally correlated case $L=20$. The analogous value for the original sequence (blue square) is also reported for a direct comparison.}
\end{figure*}

\begin{figure*}
\centering
\includegraphics[width=1.\textwidth]{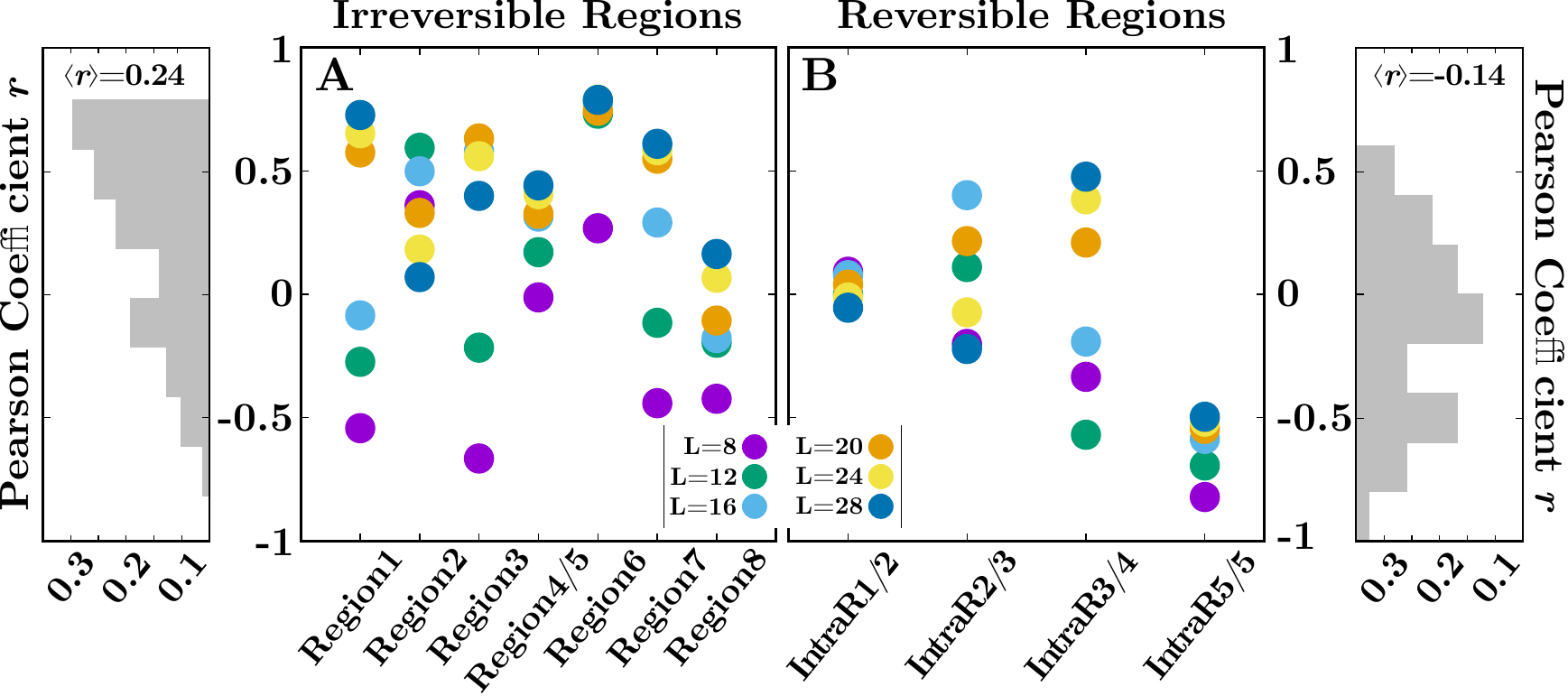}
\caption{\label{fig:RegionsPearsonCorr} Pearson correlation of the loop-BEL with the hysteresis between unfolding and refolding trajectories ($\alpha \beta = \rm UR$ in Eq.(7) of the main text) for all the irreversible \textbf{(A)} and reversible \textbf{(B)} regions for stem-loops sizes in the range [8:28]. The loop-BEL and hysteresis exhibit a higher correlation in the irreversible regions than in the reversible ones as shown by the coefficient distributions (grey bars in the leftmost end rightmost panels) that are peaked at $r^{\rm Max}_{\rm irr} \approx 0.7$ and , $r^{\rm Max}_{\rm rev} \approx -0.1$ respectively. As shown in the paper, the correlation grows with the size of the $L$-segments and is maximal for $L>16$. The $r$ value averaged over all regions and stem-loops sizes for the irreversible and reversible regions ($\langle r\rangle = 0.24$ and $\langle r\rangle = -0.14$, respectively) is also shown.}
\end{figure*}

\begin{figure*}
\centering
\includegraphics[width=1.\textwidth]{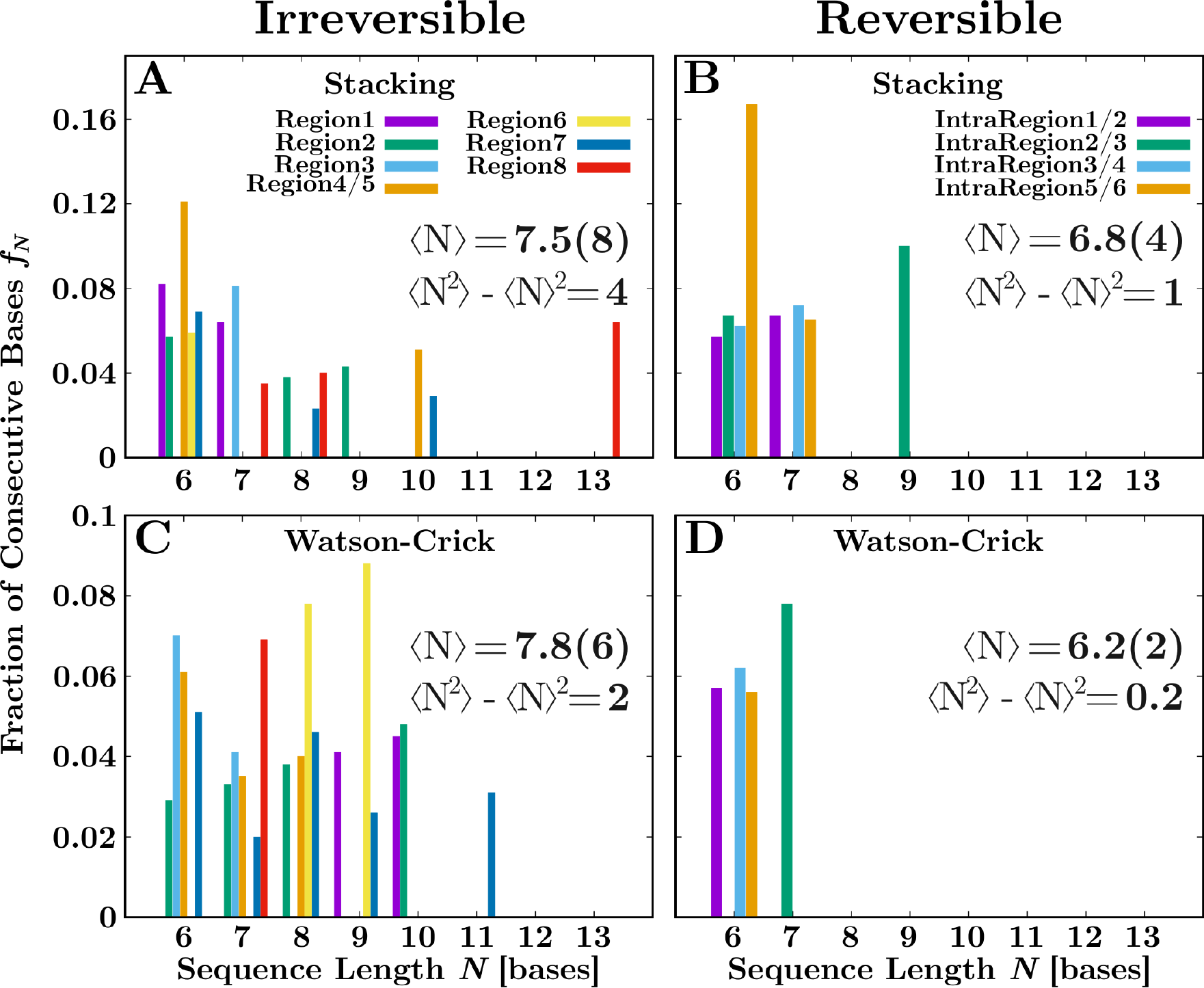}
\caption{\label{fig:SequenceAnalysis} Histogram of the results from the 2kbp RNA sequence analysis of the irreversible and reversible regions. Sequence motifs of $N \geq 6$, containing stacking (G,A) and hydrogen bonding (A,U or C,G) sequences along the unpaired strands, were analysed. \textbf{(A,B)} Stacking analysis. Fraction of consecutive stacked purines (A,G) as a function of the segment length per each irreversible (left) and reversible (right) region. \textbf{(C,D)} Watson-Crick base pairing analysis. Fraction of consecutive (A,U or C,G) as a function of the segment length per each irreversible (left) and reversible (right) region. The analysis shows both a larger average segment length $\langle N \rangle$ and variance of the segments length $\langle N^2 \rangle - \langle N \rangle^2$ in the irreversible regions. The error (in brackets) is the statistical uncertainty in the last digit.}
\end{figure*}

\begin{figure*}
\centering
\includegraphics[width=1.\textwidth]{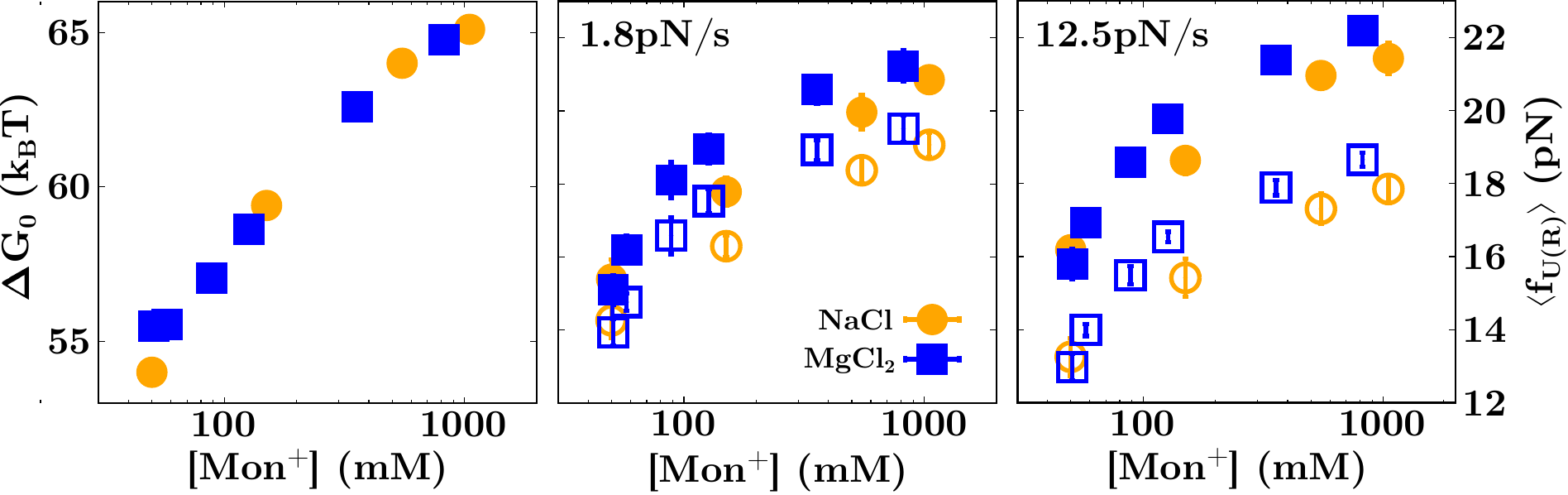}
\caption{\label{fig:CD4L4} Results from unzipping experiments of hairpin CD4 \cite{bizarro2012non} in sodium (blue squares) at 50mM, 150mM, 550mM, 1050mM and magnesium (orange circles) at 0.01mM, 0.10mM, 0.50mM, 1mM, 4mM, 10mM. \textbf{(Left)} Free energies of hybridization of the hairpin for sodium and magnesium in monovalent salt equivalents (according to the measured $77\pm 49$ equivalence rule). \textbf{(Middle)} Unzipping (full symbols) and rezipping (empty symbol) average rupture forces for sodium and magnesium in monovalent salt equivalents at 1.8pN/s pulling speed. \textbf{(Right)} Unzipping/rezipping average rupture forces for sodium and magnesium in monovalent salt equivalents at 12.5pN/s pulling speed.}
\end{figure*}

\begin{figure*}
\centering
\includegraphics[width=1.\textwidth]{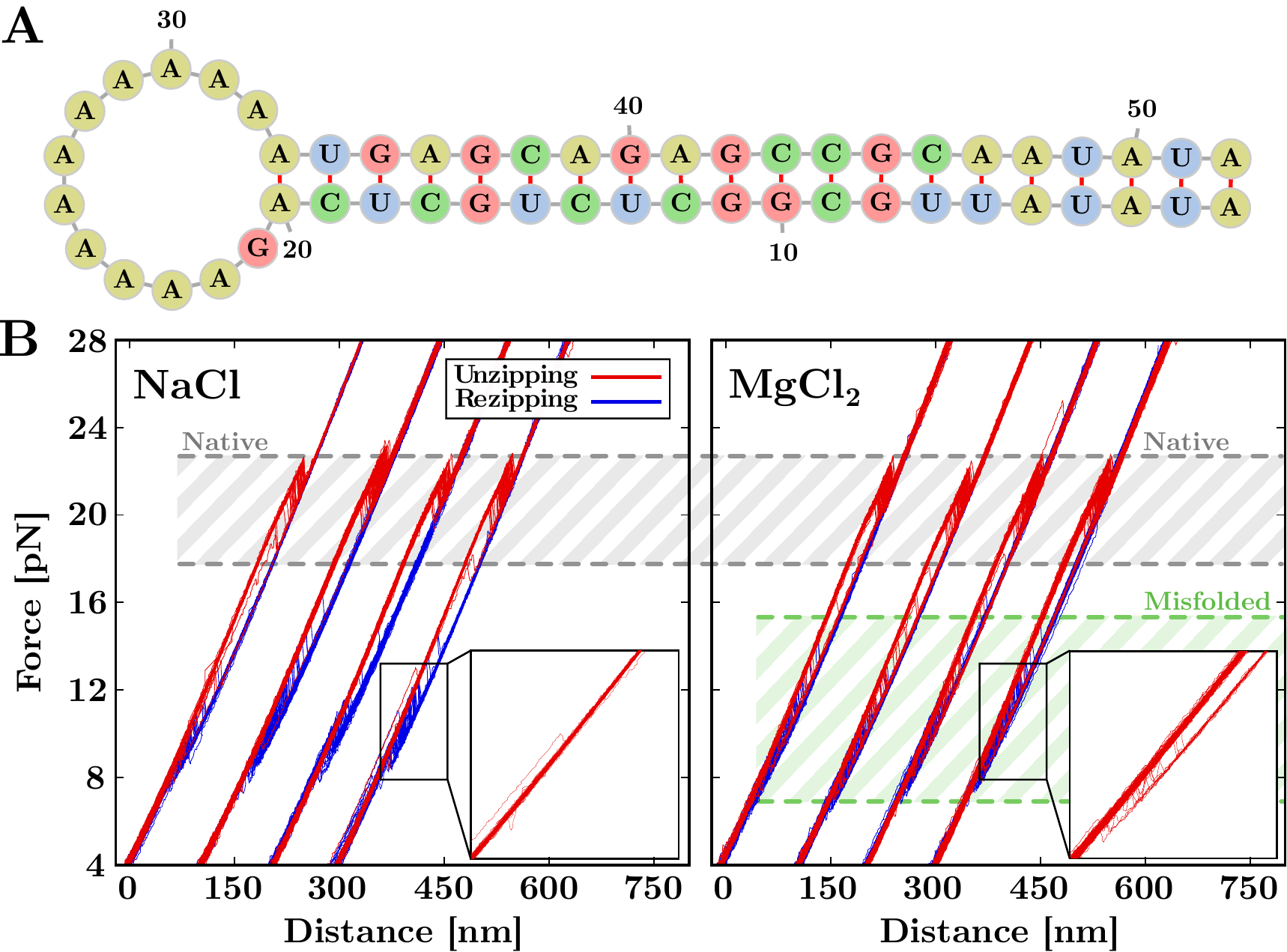}
\caption{\label{fig:HairpinA} RNA hairpin A. \textbf{(A)} Representation of the native conformation of the hairpin A stem ending with the dodecaloop GAAAAAAAAAAA. Hairpin A contains a large fraction of stacking (in the loop) and contiguous and repeated Watson-Crick base pairs (of the AU type) in the unpaired strands. \textbf{(B)} Unzipping (red) and rezipping (blue) trajectories measured for the two salt conditions 1M NaCl (left panel) and 10mM $\rm MgCl_2$ (right panel). While in sodium only the native conformation appears (gray shaded band) with a unzipping force rip at 20-22pN, experiments with magnesium reveal a misfolded state (green shaded band) occurring in the force range 7-17pN. The zooms show the unzipping FDC in the low range of forces where the misfolded state is observed in magnesium. The hairpin cartoon is based on the representation given by the Vienna RNA Web Services \cite{kerpedjiev2015forna}.}
\end{figure*}

\begin{figure*}
\centering
\includegraphics[width=1.\textwidth]{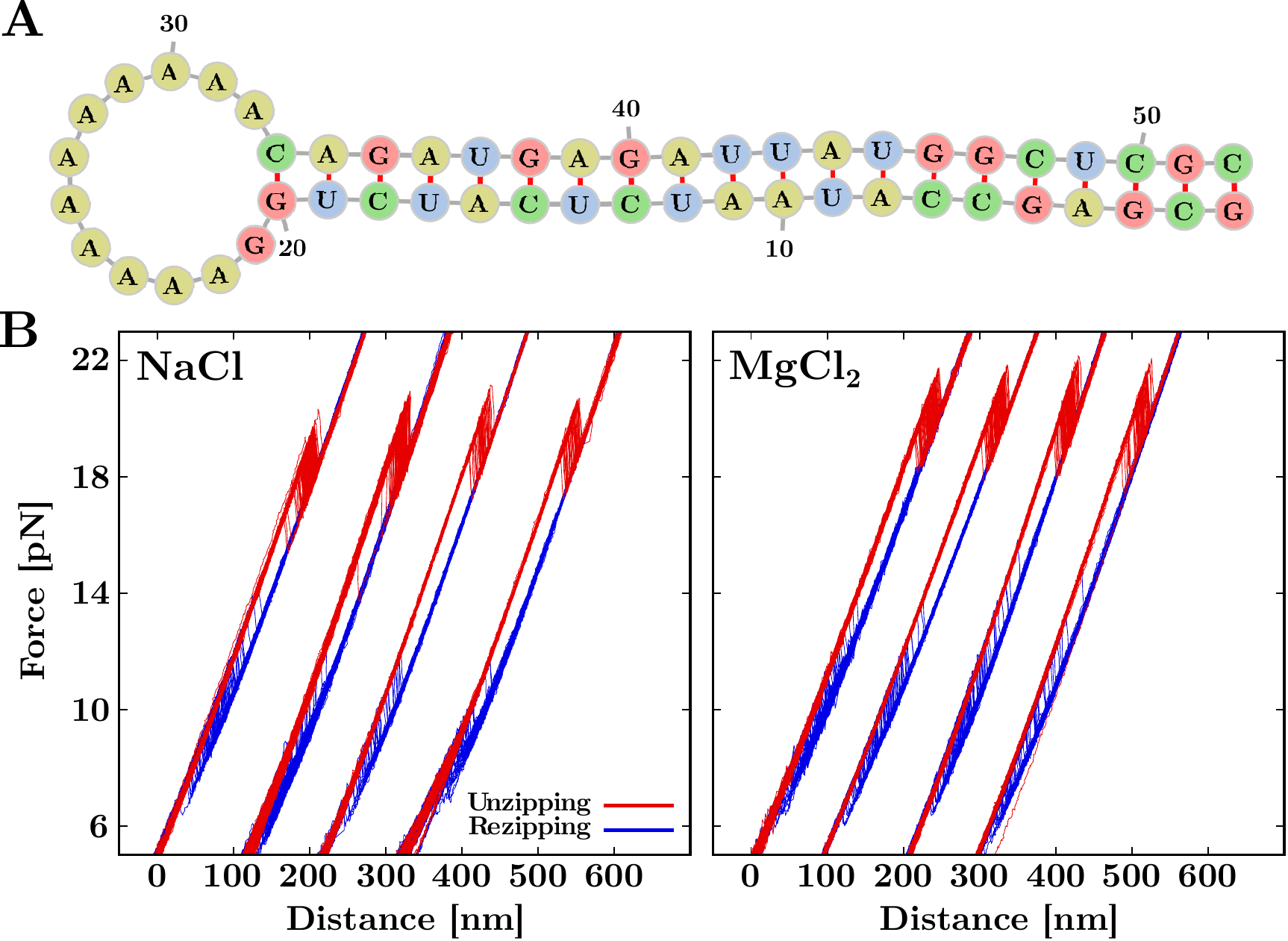}
\caption{\label{fig:HairpinB} RNA hairpin B. \textbf{(A)} Representation of the native conformation of the hairpin B, composed by the CD4 stem ending with the dodecaloop GAAAAAAAAAAA. \textbf{(B)} Unzipping experiments a the equivalent salt conditions of NaCl (300mM) and magnesium (4mM) according to the measured $77\pm49$ salt equivalence rule. The small difference in the unzipping forces in the two cases (higher for magnesium) must be attributed to the non equilibrium effect shown in Fig.\ref{fig:CD4L4}. The hairpin cartoon is based on the representation given by the Vienna RNA Web Services \cite{kerpedjiev2015forna}.}
\end{figure*}


\begin{table}
    \centering
    \caption{List of the oligonucleotides used in the synthesis of the 2027bp RNA hairpin.}
    \begin{tabular}{ m{1.9cm} l } 
     \textbf{Oligonucleotides} & \textbf{Sequence} \\ 
    \midrule
     Univ\texttt{\_}hairpin\texttt{\_}F & $5^{\prime}$-ac\textbf{gaattc}gaaaaacgcctcgagtgaag-$3^{\prime}$ \\ 
     EcoRI\texttt{\_}2.0kb\texttt{\_}R & $5^{\prime}$-ac\textbf{gaattc}ttggggtgtgtgatacgaaa-$3^{\prime}$ \\ 
     RNA1\texttt{\_}T7Forw	& $5^{\prime}$-ctaatacgactcactatagggaataaaaataggcgtatcacgag-$3^{\prime}$\\
     RNA1\texttt{\_}Rev	& $5^{\prime}$-ctcatctgtttccagatgaggggagaaaaacgcctcgagtgaag-$3^{\prime}$\\
     RNA2\texttt{\_}T7Forw	& $5^{\prime}$-ctaatacgactcactatagggagaaaaacgcctcgagtgaag-$3^{\prime}$\\
     RNA2\texttt{\_}Rev	& $5^{\prime}$-gaacatacgaaacggatgataagctgtcaaaca-$3^{\prime}$\\
     S Handle A	& $5^{\prime}$-acgaaagggcctcgtgatacgcctattttt-$3^{\prime}$\\
     S Handle B2	& $5^{\prime}$-Bio-gaacatacgaaacggatgataagctgtcaa-$3^{\prime}$\\
    \bottomrule
    \end{tabular}
    \label{tab:oligos}
\end{table}

\begin{table}
    \centering
    \caption{Occurrence of NN motifs in the RNA sequence}
    \begin{tabular}{c|c}
    \textbf{NNBP} & \textbf{Frequency (\%)} \\
    \toprule
    AA/UU & 17.1\\
    CA/GU & 14.7\\
    GA/CU & 12.7\\
    AU/UA & 9.6\\
    GU/CA & 10.0\\
    CC/GG & 8.7\\
    CG/GC & 4.1\\
    AG/UC & 11.0\\
    GC/CG & 5.6\\
    UA/AU & 6.3\\
    \end{tabular}
    \label{tab:BaseFreq}
    \caption*{\normalfont The abundance is similar for all the different NN motifs. Notice that for those motifs where it appears to be higher (AA/UU, CA/GU, GA/CU, GU/CA, AG/UC, CC/GG) there is the double degeneracy due to Watson-Crick symmetry (e.g. the fraction of AA/UU includes AA/UU and UU/AA).
}
\end{table}

%
\setlength{\tabcolsep}{1.5mm}
\begin{table}
    \centering
    \caption{\label{tab:SequenceAnalysis} Analysis of the 2kbp RNA hairpin sequence}
    \begin{tabular}{cccccccc||ccccc}
    & \multicolumn{7}{c} {\bfseries Hysteresis} & \multicolumn{3}{c} {\bfseries No Hysteresis} \\
    \addlinespace[1.mm]
    & \textbf{R1} & \textbf{R2} & \textbf{R3} & \textbf{R4} & \textbf{R5/6} & \textbf{R7} & \textbf{R8} & \textbf{IntraR1/2} & \textbf{IntraR2/3} & \textbf{IntraR3/4} & \textbf{IntraR5/6}\\
    $\mathbf{\Delta n}$ \textbf{(bases)} & 220 & 210 & 172 & 198 & 102 & 350 & 202 & 105 & 90 & 97 & 108\\
    \midrule
    \addlinespace[1.5mm]
    $\mathbf{N}$ \textbf{(bases)} & \multicolumn{11}{c} {\bfseries Number of Stacked Segments ($M_N$)} \\
    \addlinespace[1.5mm]
    \toprule
    \textbf{6} & 3 & 2 & 0 & 4 & 1 & 4 & 0 & 1 & 1 & 1 & 3\\
    \textbf{7} & 2 & 0 & 2 & 0 & 0 & 0 & 1 & 1 & 0 & 1 & 1\\
    \textbf{8} & 0 & 1 & 0 & 0 & 0 & 1 & 1 & 0 & 0 & 0 & 0\\
    \textbf{9} & 0 & 1 & 0 & 0 & 0 & 0 & 0 & 0 & 1 & 0 & 0\\
    \textbf{10} & 0 & 0 & 0 & 1 & 0 & 1 & 0 & 0 & 0 & 0 & 0\\
    \textbf{11} & 0 & 0 & 0 & 0 & 0 & 0 & 0 & 0 & 0 & 0 & 0\\
    \textbf{12} & 0 & 0 & 0 & 0 & 0 & 0 & 0 & 0 & 0 & 0 & 0\\
    \textbf{13} & 0 & 0 & 0 & 0 & 0 & 0 & 1 & 0 & 0 & 0 & 0\\
    \midrule
    \addlinespace[1.5mm]
    $\mathbf{N}$ \textbf{(bases)} & \multicolumn{11}{c} {\bfseries Number of Watson-Crick Segments ($M_N$)} \\
    \addlinespace[1.5mm]
    \toprule
    \textbf{6} & 0 & 1 & 2 & 2 & 0 & 3 & 0 & 1 & 0 & 1 & 1\\
    \textbf{7} & 0 & 1 & 1 & 1 & 0 & 1 & 2 & 0 & 1 & 0 & 0\\
    \textbf{8} & 0 & 1 & 0 & 1 & 1 & 2 & 0 & 0 & 0 & 0 & 0\\
    \textbf{9} & 1 & 0 & 0 & 0 & 1 & 1 & 0 & 0 & 0 & 0 & 0\\
    \textbf{10} & 1 & 1 & 0 & 0 & 0 & 0 & 0 & 0 & 0 & 0 & 0\\
    \textbf{11} & 0 & 0 & 0 & 0 & 0 & 1 & 0 & 0 & 0 & 0 & 0\\
    \textbf{12} & 0 & 0 & 0 & 0 & 0 & 0 & 0 & 0 & 0 & 0 & 0\\
    \textbf{13} & 0 & 0 & 0 & 0 & 0 & 0 & 0 & 0 & 0 & 0 & 0\\
    \bottomrule
    \end{tabular}
    \caption*{\normalfont Analysis of the irreversible and reversible regions along the 2kbp RNA sequence. The irreversible regions are labeled from 1 to 8 whereas the reversible ones are labeled as 'IntraR' meaning, for example, that reversible region IntraR1/R2 lies in between irreversible regions 1 and 2. The bases length $\Delta n$ of each region is also reported. \textbf{(Top)} Number of segments of $N$ consecutive stacked purines (A,G) along both strands of the stem. \textbf{(Bottom)} Number of segments of $N$ consecutive Watson-Cricks bases (A,U and C,G) along both strands of the stem.}
\end{table}
%

\FloatBarrier

\bibliography{BibliographySI}